\documentclass[%
 reprint,
 amsmath,amssymb,
 aps,
prc
]{revtex4-2}

\usepackage{graphicx}% Include figure files
\usepackage{dcolumn}% Align table columns on decimal point
\usepackage{bm}% bold math
\usepackage{hyperref}% add hypertext capabilities
\usepackage{color}
\usepackage{ulem}
\usepackage{multirow}
\usepackage{amsmath}
\usepackage{longtable}
\usepackage{CJK}
\usepackage{blindtext}
\usepackage{array}
\usepackage{tabularx}
\newcolumntype{M}[1]{>{\centering\arraybackslash}m{#1}}

\begin{document}

\begin{CJK*}{UTF8}{gbsn}

%\preprint{APS/123-QED}

%\title{Intensity of a pairing residual interaction from average single particle level densities}
\title{Determination of pairing matrix elements from average single particle level densities}

\author{Meng-Hock Koh (辜明福)}
 \email{kmhock@utm.my}
 %\homepage{https://people.utm.my/kohmenghock/}
 \affiliation{Department of Physics, Faculty of Science, Universiti Teknologi Malaysia, 
        81310 Johor Bahru, Johor, Malaysia.}%Lines break automatically or can be forced with \\
\affiliation{UTM Centre for Industrial and Applied Mathematics, 81310 Johor Bahru, Johor, Malaysia}

\author{P. Quentin}%
 \email{quentin@cenbg.in2p3.fr}
\affiliation{LP2I, UMR 5797, Universit\'{e} de Bordeaux, CNRS, F-33170, Gradignan, France
}%

\date{\today}% It is always \today, today,

\begin{abstract}
A simple and efficient method to treat nuclear pairing correlations 
within a simple Hartree-Fock--plus-BCS description is proposed and discussed.  
It relies on the fact that the intensity of pairing correlations depends crucially on level densities 
around the Fermi surface ($ \rho(e_F)$) and that any fitting of nuclear energies as functions of the nucleon numbers 
is akin of a semi-classical average, smoothing out their quantal structure. 
A particular attention has been paid to two points generally ignored in previous similar approaches. 
One is a correction advocated by M\"{o}ller and Nix [Nucl. Phys. A 536, 20 (1992)]
taking into account the fact that the data included into the fit correspond to $ \rho(e_F)$ values systematically lower than average. 
The second is due to a systematic overestimation of the proton sp level density at the Fermi surface resulting from the local Slater approximation of the Coulomb exchange terms in use in most microscopic descriptions. 
Our approach is validated by the agreement with data of corresponding calculated moments of inertia of 
well and rigidly deformed rare-earth nuclei, evaluated according to the Inglis-Belyaev ansatz 
with some crude Thouless-Valatin corrections. 
Indeed, the agreement which is found, is at least of the same quality as what results from a specific fit of the pairing intensities 
to these particular pieces of data. 
While this approach is currently limited to the very simple seniority force pairing treatment, 
it may serve as a starting point to define pairing residual interactions 
from averaged odd-even mass differences data, using merely  average sp level densities associated to calculated canonical basis.
\end{abstract}

\maketitle
\end{CJK*}

%\tableofcontents

\section{\label{sec:intro} Introduction}

We aim at discussing a simple and phenomenologically successful  approach 
to  determine the intensity of a pairing residual interaction used in a two-steps self-consistent approach 
of low energy nuclear structure as currently performed and sketched below.

A self-consistent mean field  is calculated upon using any particle-hole effective interaction 
(here the Skyrme parametrisation will be considered for the strong interaction part). 
This defines a single particle (sp) canonical basis from which pairing correlations 
are introduced either within a self-consistent Hartree-Fock (HF) approach 
(see the seminal paper \cite{NPA203}) or within  a diagonalisation 
in a  highly truncated particle-hole basis (the so-called HTDA as introduced in \cite{HTDA}).

In order to do so, one makes an appropriate choice of the residual interaction: 
seniority force (constant pairing matrix elements around the chemical potential), 
delta interaction possibly with a density dependence to enhance the surface effects \cite{SurfaceDelta}, 
Gaussian in $r$-space or separable in $p$-space \cite{GaussPsep}. 
Here, as a first step, to demonstrate the validity and performance of the 
general approach we will consider the simple seniority force ansatz.

Bohr, Mottelson and Pines \cite{BMP} have pointed out, in particular, 
two nuclear spectroscopic properties strongly contingent upon pairing correlations 
(odd-even mass differences $\delta E$ and moments of inertia $\mathcal{J}$ (MoI), i.e., 
in practice, the first 2$^{+}$ energies in well and rigidly deformed nuclei). 
Both are {\it{a priori}} accessible to a theoretical description within the above defined self-consistent approaches.

In a previous paper \cite{Hafiza2019}, making two independent fits of these two properties in the rare earth region, it has been shown that one has obtained very similar values for the parameters of the seniority force matrix elements. This provides hints that: a) as expected, pairing correlations are indeed (all things being kept similar) the main factor to yield precise values of $\delta E$ as well as $\mathcal{J}$, and b) that such a simplified approach, also followed here, was suited globally to describe these properties.

However, in both approaches the quality of the agreement with data was  locally (i.e. around a given nucleus) contingent upon a perfect reproduction of the ordering and fine distribution of the sp energies. But a fit corresponds in its very principle to a reproduction of some properties on the average. The above quoted success was obtained  upon considering reasonably sized samples so one could conclude that on the average the regime of pairing correlations was correctly adjusted. 

Taking stock of this remark, we have chosen here to perform an estimate of the pairing strengths 
using a tool {\it{a priori}} less sensitive to local sp spectrum deficiencies.
The question we ask ourselves here is whether in doing so, 
we would obtain a reproduction of MoI at least as good as what has been obtained by a direct fit of these moments (see Ref. \cite{Hafiza2019}).

In practice we used the so-called uniform gap method (\cite{FunnyHill}, see also Ref.~\cite{RingSchuck})
providing a value of the average matrix elements (around the Fermi energy) of the pairing residual interaction upon adopting an adequate smooth parametrisation of the neutron and proton pairing gaps $\Delta (N,Z)$.

Some special attention has been paid to two important points of a different nature.

The first issue is of a physical nature. Some analytical forms of the $\Delta (N,Z)$ function have been proposed by Jensen \textit{et al.} \cite{Jensen1984} 
and Madland and Nix \cite{Madland1988}, but M\"{o}ller and Nix \cite{Moller1992} 
have pointed out a systematic bias of these approaches 
related with the lower than average character of the sp level density at equilibrium deformations. 
They proposed a phenomenological correction which we will adopt in this paper. 

The second point concerns an approximation made in most self-consistent calculations (including most of ours) for the sake of numerical easiness. 
In previous studies (e.g. in Table II of \cite{Hafiza2019}) one has noticed that the treatment of pairing correlations 
was significantly more successful for neutrons than it was for protons. 
It has been hinted that it was related to an approximate treatment of the non-local Fock term of the Coulomb mean field due to Slater \cite{Slater}. 
As discussed in Sub-Section~\ref{sec: proton effective pairing gaps}, 
this approximation systematically overestimates the sp level density near the Fermi level \cite{Titin,Skalski,Bloas}. 
A proper account of this spurious enhancement was thus in order and an appropriate correction to the proton 
sp level density has been implemented upon comparing 
the effect of the Coulomb exchange terms issued from approximate and exact calculations.

Calculations have been performed for a sample of $19$ lanthanide nuclei
supplemented by three isotopes of Hafnium and one isotope of Tungsten
(hereafter loosely refered to as rare earth nuclei) and 22 actinide nuclei. 
They are listed in Table~\ref{tab:moi and fitted pairing results}. 
These nuclei are well ($\beta_{20}$ values in the  $0.2 - 0.3$ range with usual notation) and rigidly deformed.
The latter property is ascertained by a ratio of their
first $2^{+}$ and $4^{+}$ levels $R_{42} \ge 3.290$ as displayed
in Table~\ref{tab:moi and fitted pairing results} (energy data taken from the compilation of Ref. \cite{NNDC}).

\begin{table*}[]
\caption{The ratio of the first $4^+$ over $2^+$ energy $R_{42}$, 
    the estimated neutron $V_n$ and proton $V_p$ pairing matrix elements (in MeV),
    the calculated MoI (in $\hbar^2/\mbox{MeV}$), including a Thouless-Valatin corrective factor of $\alpha = 1.32$, obtained with
    the SIII, SkM* and SLy4 parametrisations and
    the experimental MoI extracted from the first $2^+$ energy, $\mathcal{J}_{exp}$,
    for a nucleus with $Z$ protons and $N$ neutrons.
    Actinide nuclei removed from the MoI r.m.s analyses presented in Table~\ref{tab:rms moi} are marked with dashed lines.
    \label{tab:moi and fitted pairing results}
    }
\begin{ruledtabular}
\begin{tabular}{*{14}c}
\multirow{2}{*}{Z}	&	\multirow{2}{*}{N}	&	\multirow{2}{*}{A}	&	\multirow{2}{*}{$R_{42}$}&   
    \multicolumn{3}{c}{SIII}&   \multicolumn{3}{c}{SkM*}&   \multicolumn{3}{c}{SLy4}&    \multirow{2}{*}{$\mathcal{J}_{exp}$}\\
    \cline{5-7} \cline{8-10} \cline{11-13}
&&&&    $V_n$	&	$V_p$	&	$\mathcal{J}_{TV}$	&	$V_n$	&	$V_p$	&	$\mathcal{J}_{TV}$	&	
        $V_n$	&	$V_p$	&	$\mathcal{J}_{TV}$	&		\\
\colrule
62	&	94	&	156	&	3.290	&	0.1901	&	0.2529	&	37.742	&	0.1769	&	0.2450	&	39.438	&	0.2013	&	0.2537	&	38.221	&	39.531	\\
62	&	96	&	158	&	3.301	&	0.1862	&	0.2517	&	38.068	&	0.1727	&	0.2440	&	42.063	&	0.1964	&	0.2527	&	37.148	&	41.209	\\
62	&	98	&	160	&	3.292	&	0.1818	&	0.2510	&	40.765	&	0.1683	&	0.2435	&	40.693	&	0.1916	&	0.2518	&	39.215	&	42.373	\\
64	&	96	&	160	&	3.302	&	0.1865	&	0.2454	&	36.537	&	0.1735	&	0.2383	&	39.943	&	0.1970	&	0.2467	&	35.192	&	39.860	\\
64	&	98	&	162	&	3.302	&	0.1822	&	0.2442	&	40.663	&	0.1694	&	0.2377	&	38.840	&	0.1925	&	0.2453	&	38.258	&	41.899	\\
64	&	100	&	164	&	3.295	&	0.1776	&	0.2430	&	40.657	&	0.1651	&	0.2370	&	38.136	&	0.1880	&	0.2445	&	42.849	&	40.944	\\
64	&	102	&	166	&	3.297	&	0.1738	&	0.2422	&	42.775	&	0.1612	&	0.2366	&	39.549	&	0.1843	&	0.2437	&	37.230	&	42.857	\\
66	&	96	&	162	&	3.294	&	0.1867	&	0.2389	&	35.283	&	0.1742	&	0.2308	&	38.742	&	0.1974	&	0.2393	&	35.302	&	37.193	\\
66	&	98	&	164	&	3.301	&	0.1825	&	0.2379	&	39.453	&	0.1701	&	0.2302	&	38.577	&	0.1931	&	0.2381	&	39.004	&	40.876	\\
66	&	100	&	166	&	3.310	&	0.1783	&	0.2373	&	38.850	&	0.1645	&	0.2294	&	38.936	&	0.1888	&	0.2372	&	43.868	&	39.171	\\
66	&	102	&	168	&	3.313	&	0.1743	&	0.2367	&	40.901	&	0.1564	&	0.2290	&	42.649	&	0.1847	&	0.2359	&	38.845	&	40.021	\\
68	&	100	&	168	&	3.309	&	0.1797	&	0.2318	&	36.297	&	0.1669	&	0.2224	&	37.682	&	0.1893	&	0.2319	&	41.603	&	37.592	\\
68	&	102	&	170	&	3.310	&	0.1749	&	0.2307	&	39.076	&	0.1631	&	0.2217	&	39.339	&	0.1854	&	0.2302	&	37.400	&	38.173	\\
68	&	104	&	172	&	3.314	&	0.1712	&	0.2296	&	35.341	&	0.1605	&	0.2209	&	39.610	&	0.1796	&	0.2287	&	36.688	&	38.961	\\
70	&	100	&	170	&	3.293	&	0.1791	&	0.2261	&	35.754	&	0.1674	&	0.2167	&	38.655	&	0.1895	&	0.2266	&	40.594	&	35.606	\\
70	&	102	&	172	&	3.305	&	0.1753	&	0.2246	&	37.865	&	0.1637	&	0.2154	&	41.260	&	0.1859	&	0.2244	&	37.624	&	38.099	\\
70	&	104	&	174	&	3.310	&	0.1717	&	0.2227	&	35.636	&	0.1602	&	0.2135	&	43.698	&	0.1823	&	0.2221	&	37.465	&	39.231	\\
70	&	106	&	176	&	3.310	&	0.1681	&	0.2215	&	34.882	&	0.1567	&	0.2120	&	42.205	&	0.1786	&	0.2202	&	38.821	&	36.525	\\
70	&	108	&	178	&	3.310	&	0.1646	&	0.2204	&	37.397	&	0.1533	&	0.2109	&	42.128	&	0.1756	&	0.2185	&	38.506	&	35.714	\\
72	&	106	&	178	&	3.291	&	0.1686	&	0.2164	&	31.988	&	0.1573	&	0.2108	&	33.572	&	0.1789	&	0.2188	&	33.180	&	32.196	\\
72	&	108	&	180	&	3.307	&	0.1652	&	0.2154	&	33.615	&	0.1539	&	0.2097	&	31.557	&	0.1753	&	0.2175	&	31.464	&	32.146	\\
72	&	110	&	182	&	3.295	&	0.1617	&	0.2144	&	31.436	&	0.1506	&	0.2089	&	31.185	&	0.1725	&	0.2165	&	31.184	&	30.678	\\
74	&	108	&	182	&	3.291	&	0.1653	&	0.2112	&	30.384	&	0.1543	&	0.2061	&	27.708	&	0.1753	&	0.2141	&	27.755	&	29.968	\\
\colrule
90	&	144	&	234	&	3.291	&	0.1216	&	0.1744	&	57.708	&	0.1131	&	0.1671	&	65.546	&	0.1297	&	0.1715	&	74.584	&	60.545	\\
92	&	140	&	232	&	3.291	&	0.1254	&	0.1632	&	63.879	&	0.1172	&	0.1628	&	-	&	0.1340	&	0.1678	&	76.273	&	63.061	\\
92	&	142	&	234	&	3.296	&	0.1235	&	0.1698	&	-	&	0.1153	&	0.1630	&	-	&	0.1291	&	0.1677	&	82.008	&	68.969	\\
92	&	144	&	236	&	3.304	&	0.1217	&	0.1700	&	63.052	&	0.1135	&	0.1634	&	-	&	0.1300	&	0.1678	&	84.126	&	66.307	\\
92	&	146	&	238	&	3.303	&	0.1203	&	0.1701	&	59.663	&	0.1117	&	0.1639	&	-	&	0.1281	&	0.1682	&	71.416	&	66.791	\\
92	&	148	&	240	&	3.347	&	0.1181	&	0.1704	&	-	&	0.1099	&	0.1642	&	-	&	0.1262	&	0.1691	&	63.613	&	66.667	\\
94	&	142	&	236	&	3.304	&	0.1237	&	0.1654	&	-	&	0.1156	&	0.1599	&	71.615	&	0.1291	&	0.1593	&	-	&	67.219	\\
94	&	144	&	238	&	3.312	&	0.1219	&	0.1654	&	67.935	&	0.1138	&	0.1601	&	70.205	&	0.1250	&	0.1645	&	-	&	68.081	\\
94	&	146	&	240	&	3.309	&	0.1201	&	0.1656	&	65.827	&	0.1120	&	0.1602	&	73.433	&	0.1284	&	0.1647	&	71.278	&	70.054	\\
94	&	148	&	242	&	3.307	&	0.1183	&	0.1703	&	61.921	&	0.1103	&	0.1606	&	-	&	0.1265	&	0.1650	&	66.699	&	67.355	\\
94	&	150	&	244	&	3.391	&	0.1167	&	0.1663	&	63.423	&	0.1086	&	0.1609	&	64.584	&	0.1248	&	0.1654	&	73.284	&	67.873	\\
94	&	152	&	246	&	3.308	&	0.1150	&	0.1669	&	-	&	0.1075	&	0.1612	&	59.134	&	0.1233	&	0.1656	&	59.224	&	64.240	\\
96	&	146	&	242	&	3.252	&	0.1202	&	0.1622	&	-	&	0.1123	&	0.1565	&	-	&	0.1290	&	0.1611	&	-	&	71.208	\\
96	&	148	&	244	&	3.314	&	0.1185	&	0.1626	&	66.225	&	0.1061	&	0.1566	&	-	&	0.1267	&	0.1610	&	70.430	&	69.837	\\
96	&	150	&	246	&	3.313	&	0.1169	&	0.1630	&	65.332	&	0.1091	&	0.1569	&	68.096	&	0.1250	&	0.1612	&	-	&	70.008	\\
96	&	152	&	248	&	3.313	&	0.1152	&	0.1635	&	-	&	0.1073	&	0.1571	&	64.122	&	0.1197	&	0.1614	&	65.674	&	69.124	\\
98	&	150	&	248	&	3.318	&	0.1170	&	0.1597	&	63.764	&	0.1093	&	0.1532	&	71.893	&	0.1251	&	0.1592	&	-	&	72.237	\\
98	&	152	&	250	&	3.321	&	0.1153	&	0.1599	&	-	&	0.1076	&	0.1534	&	68.774	&	0.1234	&	0.1578	&	67.810	&	70.223	\\
98	&	154	&	252	&	3.319	&	0.1138	&	0.1640	&	-	&	0.1060	&	0.1537	&	62.400	&	0.1216	&	0.1579	&	61.026	&	65.617	\\
100	&	154	&	254	&	3.319	&	0.1139	&	0.1560	&	-	&	0.1063	&	0.1499	&	64.823	&	0.1218	&	0.1546	&	62.273	&	66.679	\\
100	&	156	&	256	&	3.317	&	0.1098	&	0.1560	&	-	&	0.1047	&	0.1502	&	59.527	&	0.1201	&	0.1545	&	61.250	&	62.344	\\
102	&	150	&	252	&	3.310	&	0.1152	&	0.1520	&	67.468	&	0.1097	&	0.1469	&	-	&	0.1252	&	0.1521	&	-	&	64.655	\\
%102	&	152	&	254	&	3.285	&	0.1155	&	0.1519	&	-	&	0.1081	&	0.1467	&	-	&	0.1235	&	0.1515	&	-	&	67.873	\\
\end{tabular}
\end{ruledtabular}
\end{table*}

The paper is organised as follows. The general fitting approach is presented in Section~\ref{sec: approach}. 
The extraction of averaged sp level density is performed \textit{\`{a} la} Strutinsky \cite{FunnyHill} 
from our microscopic calculations for a sample of well and rigidly deformed nuclei. 
The uniform gap method is used to extract the matrix element of the pairing residual interaction 
upon using the M\"{o}ller-Nix ansatz for average $\Delta (N,Z)$ values. 
The definition of effective average gaps to be fitted as well as the correction in the proton case 
for the approximation made on the Fock Coulomb terms are also discussed there. 
Some technical details are briefly presented in Section~\ref{sec:technical details}. 
They include the choice of the sample of deformed nuclei in the two considered regions of heavy deformed nuclei (around rare-earth and actinide elements) and the specific choice made for the Strutinsky averaging of the sp level density. 
Our results obtained with three
parametrisations of the Skyrme interaction, 
namely SIII \cite{SIII},
SkM* \cite{SkMs} and SLy4 \cite{SLy4}, for the moments of inertia
are presented in Section~\ref{sec:results}. 
Finally, Section~\ref{sec: conclusion} summarizes the main conclusions of our work.

\section{The approach}
\label{sec: approach}
\subsection{Overview of the approach}
\label{sec: overview of the approach}

Assuming that we know the smooth behaviour of the average neutron and proton pairing gaps with $N$ and $Z$, 
we devise here an approach to get the corresponding smooth evolution of pairing matrix elements $V_q$ (where $q$ stands for the charge states)
averaged over a given sp valence space being contained 
in the $[\lambda_q - \Omega, \lambda_q + \Omega]$ range where $\lambda_q$ is the Fermi energy to be defined later, 
while the above energy interval (spanning a $2 \Omega$ energy range) 
characterizes the domain of sp states (of the canonical basis) active in the BCS treatment. 
%In this work we will take  $\Omega = 6$ MeV. 

From the exact sp level density $\rho(e)$ as a function (rigorously speaking distribution) of the energy $e$, 
we define a semi-classical sp level density function $\tilde{\rho}(e)$ obtained in practice by a Strutinsky averaging in $e$ \cite{FunnyHill}. 
We recall here the close connection of the Strutinsky energy averaging with a semi-classical averaging \textit{\`{a} la} Wigner-Kirkwood 
(see Ref. \cite{Jennings1975}). 

As stated in the introduction, we restrict ourselves in this work to constant pairing matrix elements $V_q$ for each charge state $q$
within an energy range of  $ 2 \Omega$ centered around averaged Fermi energies $\tilde{\lambda}_q$ defined below.
Limiting the interval of sp states active in the BCS variational process makes the value of $V_q$ dependent on the value of $\Omega$, as well known. 
Here consistently we will take $\Omega = 6$ MeV.

The matrix elements $V_q$ are determined in terms of
the averaged sp level densities $\tilde{\rho}_q (e)$ and the 
average pairing gaps $\tilde{\Delta}_q$ through the following gap equation:
%For each charge state $q$, the matrix elements $V_q$ of the pairing residual interaction
%are given as functions of the average gaps $\tilde{\Delta}_q$ and the Fermi energies $\lambda_q$ by the following BCS gap equation
\begin{equation}
\frac{1}{V_q}  = \int_{\tilde{\lambda}_q - \Omega}^{\tilde{\lambda}_q + \Omega} 
\frac{\tilde{\rho}_q(e)}{\sqrt{(e - \tilde{\lambda}_q)^2 +{\tilde{\Delta}}_q^2}} de.
\label{eq: integration of ave density for pairing matrix element}
\end{equation}

The Fermi energies $\tilde{\lambda}_q$ are defined from the average density ${\tilde{\rho}}_q(e)$ for a total fermion number $N_q$ such that
\begin{equation}
N_q  = \int_{-\infty}^{\tilde{\lambda}_q} \tilde{\rho}_q(e) de.
\end{equation}

A frequently used approximation of the above gap equation consists in assuming that the variation of  ${\tilde{\rho}}_q(e)$ 
is small enough within the $[\tilde{\lambda}_q - \Omega, \tilde{\lambda}_q + \Omega]$ 
sp energy interval so that one can replace it by a constant, namely its Fermi energy value $ {\tilde{\rho}}_q(\lambda_q) $. 
Upon performing the integral one then obtains a closed form formula for $V_q$ as
\begin{equation}
\frac{1}{V_q}  = \frac{1}{2}{\tilde{\rho}}_q(\lambda_q) sinh^{-1} \Big(\frac{\Omega}{{\tilde\Delta}}_q\Big).
\label{eq: asinh function for pairing matrix element}
\end{equation}

\begin{figure}
    \centering
    \includegraphics[width = 0.5\textwidth]{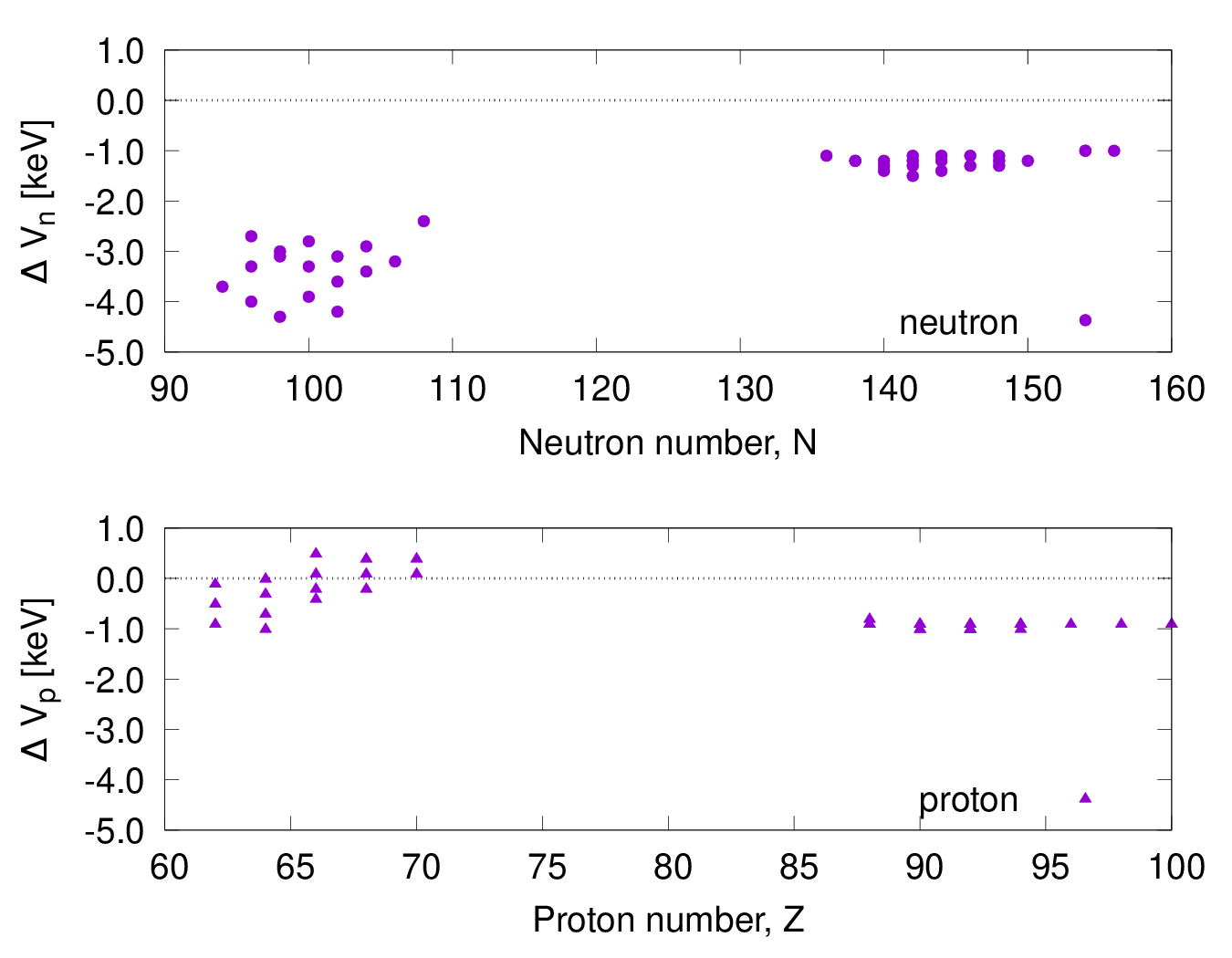}
    \caption{Difference in pairing matrix elements $\Delta v_q$
    between values obtained using a full integration over the pairing window $\Omega$ 
    and those obtained using the asinh function given in equation (\ref{eq: asinh function for pairing matrix element})}
    \label{fig:difference pairing matrix element exact vs asinh}
\end{figure}

This approximation is shown to be indeed rather good as displayed in Figure~\ref{fig:difference pairing matrix element exact vs asinh}
where we have plotted the difference in the pairing matrix elements obtained using 
equation (\ref{eq: integration of ave density for pairing matrix element}) and (\ref{eq: asinh function for pairing matrix element})
using an integration interval defined by  $\Omega = 6$ MeV.
The difference in the proton pairing matrix elements between the two equations is mostly localized between $\pm 1$ keV
while the largest difference for neutrons in absolute value is $4.3$ keV.
%Yet since we need to compute the smooth densities ${\tilde{\rho}}_q(e)$ anyway, we do not need to have recourse to it.

While we have shown {\it{en passant}} that the approximate equation (\ref{eq: asinh function for pairing matrix element}) is a rather good approximation,
we have however resorted to making a full integration using equation (\ref{eq: integration of ave density for pairing matrix element}) for our calculations.

At the end of this process, we will then have for each charge state an average matrix element 
of the pairing residual interaction $V_q(N,Z)$ as a function of $N$ and $Z$.

\subsection{Determination of the average single-particle level density}

The crucial part is then to determine the average level density for a given set of sp levels.
We have computed the average level density using the equation \cite{FunnyHill,NPA207}
\begin{equation}
    \tilde{\rho}_q (e) = \frac{1}{\gamma} \int_{-\infty}^{\infty} \rho(e') f\Big( \frac{e' - e}{\gamma}\Big) de'.
    \label{eq:average level density}
\end{equation}
The so-called curvature corrections (as discussed in Refs. \cite{FunnyHill,NPA207}) are taken care of by the $f(x)$ term defined as 
\begin{equation}
    f(x) = P(x) \: w(x)
\end{equation} 
where $P(x)$ is a polynomial of degree $2M$ in $x$ defined in terms of  generalized Laguerre polynomials $L_{M}^{(\alpha)}$ of the form
\begin{equation}
    P(x) = L_M^{1/2}(x^2) = \sum_{n = 0}^M a_{2n} \: x^{2n}.
    \label{eq: polynomial eq}
\end{equation}
with coefficients $a_{2n}$ given in Table~\ref{tab:polynomial coefficients}
and $w(x)$ being a weightage of Gaussian type defined by
\begin{equation}
    w(x) = \frac{1}{\sqrt{\pi}} e^{-x^2}.
\end{equation}

\begin{table}
    \centering
    \begin{tabular}{m{1.2cm} m{1.2cm} m{1.2cm} m{1.2cm} m{1.2cm} m{1.2cm}}
    \hline
      M&     $a_0$&   $a_2$&    $a_4$&    $a_6$&    $a_8$ \\
      \hline
      0&       1&    --&     --&     --&    --  \\
      1&     3/2&    -1&     --&     --&    --  \\
      2&     15/8&   -5/2&   1/2&    --&    --  \\
      3&     35/16&  -35/8&  7/4&    -1/6&  --  \\
      4&    945/384&    -315/48&    63/16&  -9/12&  1/24    \\
      \hline
    \end{tabular}
    \caption{Five lowest generalized Laguerre polynomial coefficients entering equation~(\ref{eq: polynomial eq}).}
    \label{tab:polynomial coefficients}
\end{table}

The value of the smoothing width $\gamma$ appearing in equation~(\ref{eq:average level density}) is crucial to define a correct
energy window for the discrete sp levels to be considered in the integration of equation~(\ref{eq:average level density}). 
This, as discussed e.g. in Refs. \cite{FunnyHill,NPA207}, is generally defined by fulfilling the so-called {\textit{plateau condition}} 
ensuring that the shell effect energy is almost constant as a function of $\gamma$. 

In practice, we chose a value of $\gamma$ (in MeV) according to the total nucleon number $A$ as
\begin{equation}
    \gamma = \beta \: \frac{41}{A^{1/3}} [MeV]
\end{equation}
using the standard (see Ref.~\cite{Moszkowski}) formula for the energy spacing between major oscillator shells.
The specific value of the above constant $\beta$ and the polynomial order $2 M$ in $x$ 
of $P(x)$ will be discussed in Sub-Section~\ref{sec:choice of coefficients}.

\subsection{Empirical vs effective-pairing gaps
\label{sec: effective pairing gaps}}
The next crucial ingredients to the approach 
are the pairing gaps extracted from the data on $\delta E$,
entering equation~(\ref{eq: integration of ave density for pairing matrix element})
or (\ref{eq: asinh function for pairing matrix element}).
At this point, it is important to differentiate between empirical pairing gaps as used e.g. in Refs. \cite{Jensen1984,Madland1988}
and those to be employed in any fitting approach such as ours.
The latter are referred to as effective interaction pairing gaps in \cite{Moller1992}. 

As discussed in \cite{Moller1992}, these effective gap values should take into account a bias due to the shell effects 
in the sp level density at equilibrium deformation making it systematically lower than its average value. 
Therefore the experimental pairing gaps should not be used as such to determine average pairing properties. 
Their values should be quenched since upon not doing that a fit of the residual interaction on them 
would lead to an overestimation of the pairing correlations in actual Hartree-Fock--plus--BCS (HF+BCS) calculations. 

Herein, we use the effective-interaction average pairing gaps, phenomenologically determined in Ref. \cite{Moller1992} to be:
\begin{equation}
    \tilde{\Delta}_{q} = \frac{r B_s}{N_q^{1/3}}
    \label{eq: Moller Nix pairing gap}
\end{equation}
where $B_s$ is set to $1$ and $r = 4.8$ MeV \cite{Moller1992}.

\subsection{Effective pairing gaps in the proton case
\label{sec: proton effective pairing gaps}}

In the case of protons, the above value of $r$ yields slightly too high BCS pairing gaps since the above formula corresponds 
supposedly to an exact treatment of the Coulomb interaction
This is, however, not the case in most mean-field calculations. In order to avoid considering non-local mean fields,
the Coulomb-exchange contribution is usually accounted for \textit{\`{a} la} Slater, i.e. upon using the infinite nuclear matter Pauli correlation function \cite{Slater}. 
As found long ago \cite{Titin} and confirmed later \cite{Skalski,Bloas} the Slater approximation in use, 
systematically overestimates the sp level density near the Fermi level with respect to the corresponding exact treatment. 

In order to correct for this systematic spurious trend modifying significantly the level density, we have proceeded as follows.
We have considered for a given nucleus, two sp proton spectra obtained within the Skyrme HF+BCS framework, 
corresponding to an exact treatment and the Slater approximation of the Coulomb exchange terms. 
The exact calculation of the Coulomb energy matrix elements has been performed according 
to the method developed in Refs. \cite{Titin} and \cite{Bloas}.
This has yielded a corresponding ratio $R_{p}$ of the BCS gaps:
\begin{equation}
    R_{p} = \frac{\Delta_{p}^{exact}}{\Delta_{p}^{Slater}}.
    \label{eq: R_p}
\end{equation}

We take for granted the average effective gap of Ref.~\cite{Moller1992} (see equation (47)) with $r = 4.8$ MeV as reproducing
adequately effective average pairing matrix elements. 
For a given well and rigidly deformed nucleus and a given interaction, 
we perform calculations with the Slater approximation using the M\"{o}ller-Nix value of $r$
and \textit{reasonable} values (as defined below) of the average pairing matrix element for instance as
(see the discussion of such a choice in Subsection~\ref{sec:Generation of sp levels})

It is well known that the sp spectra are almost unaffected by a variation of $V_q^0$ in quite a large range of
values of around values such as those displayed above in eq.~(\ref{eq: seniority force 1}).
This is demonstrated at least for the average sp level densities at the Fermi surface in Table~\ref{tab:compare G_q when making fit}.
For two nuclei in their deformed ground states, $^{176}$Yb and $^{240}$Pu, 
we have calculated $\tilde{\rho}_q (\lambda_q)$ for three values of the intensity parameter
$G_q = 16, 19, 22$ MeV. 
They vary for both nuclei and for all values of $G_q$ no more than $0.1 \%$.
Therefore, the resulting values of the average pairing matrix elements $V_q^1$ contingent merely upon the
sp spectra should seemingly not depend on the arbitrary chosen values of $V_q^0$. 
This is clearly so for the neutron matrix elements $V_n$ and not really the case for the 
proton matrix elements $V_p$, for reasons which will be discussed now.

\begin{equation}
    V_q^0 = \frac{G_q}{11 + N_q} \;\;\; \mbox{with} \; \forall_q, \; G_q = 19 \; \mbox{MeV}.
    \label{eq: seniority force 1}
\end{equation}

\begin{table}
    \caption{The average Fermi level densities for neutrons and protons (in MeV$^{-1}$)
            are reported in columns 4 and 5, respectively,
            for the $^{176}$Yb and $^{240}$Pu nuclei
            in the ground-state deformation characterized by its quadrupole moment $Q_{20}$ (in barns).
            Columns 6 and 7 give the corresponding fitted  values (in MeV) 
            of the average matrix elements $V_q$ (with $M = 2$, $\beta = 1.2$).
            These values are generated from HF+BCS calculations using different $G_q$ values (equal for neutrons and protons) as reported in column 2.
            }
            
    \label{tab:compare G_q when making fit}
%    \centering
%    \begin{tabular}{m{1.2cm} m{1.2cm} m{1.2cm} m{1.2cm} m{1.2cm} m{1.2cm}}
\begin{ruledtabular}
    \begin{tabular}{*{7}c}
        %\hline
          Nucleus&   $G_q$&  $Q_{20}$& ${\tilde{\rho}}_n(\lambda_n)$& $ {\tilde{\rho}}_p(\lambda_p)$&	  $V_n$&    $V_p$  \\
         \hline
        \multirow{3}{*}{Yb-176} & 16&	 19.13&   4.709&	3.619&  0.1681&	0.2151 \\
                                & 19&	 18.73&   4.710&	3.622&  0.1681&	0.2215 \\
                                & 22&	 18.34&   4.717&	3.625&  0.1680&	0.2300  \\
        \hline
        \multirow{3}{*}{Pu-240}&  16&   28.72&    6.388&  4.736&  0.1201&   0.1602  \\
                                & 19&   28.26&    6.386&  4.735&  0.1201&   0.1656 \\
                                & 22&   27.77&    6.385&  4.735&  0.1201&   0.1733\\
    \end{tabular}
\end{ruledtabular}
\end{table}

It is not as simple as to perform calculations with these $V_p^0$ values taking
exactly into account the Coulomb exchange calculations to get $\Delta_p^{exact}$ and 
then determine $R_p$ as defined in eq~(\ref{eq: R_p}) to correct for the M\"{o}ller-Nix parameter. 
It turns out that the value of $R_p$ depends on the chosen value of $V_p^0$ or in effect
on the degree of pairing correlations as a monotonically increasing function. 
This could be expected since the more pairing correlations are present, 
the closer to a smoothed-out level density distribution one would get, 
making it more and more closer to the one present in infinite nuclear matter corresponding to the
Slater approximation.

%\textcolor{red}{\sout{Let us assume, as it will be discussed in Appendix~\ref{sec:appendix a}, 
%that we have at our disposal a universal formula relating $R_p$
%with a quantity representing the degree of pairing correlations.}}
To quantify the degree of pairing correlations, 
we consider the so-called pair condensation energy $E_{cond}^p$ for protons
(the absolute value of the part of the HF+BCS
energy which involves the abnormal density), given by
\begin{equation}
    E_{cond}^p = \frac{\Delta_p^2}{V_p}.
    \label{eq: condensation energy}
\end{equation}

To determine the variation of $R_p$ as a function of $E_{cond}^p$, we consider a sample 
of 19 nuclei in the case of SIII (17 nuclei for SLy4 and SkM*)
with respect to the 45 nuclei listed in Table~\ref{tab:moi and fitted pairing results}.
Specifically, we have excluded nuclei exhibiting
large sp energy gaps leading to artifically low BCS pairing gap, 
in view of the well-known deficiency of the BCS approximation in such weak pairing
regimes (see e.g. Ref~\cite{Zheng_1992}).

For these nuclei, we performed both exact Coulomb and Slater approximation calculations
using the pairing matrix elements listed in Table~\ref{tab:slater exact ratio}.
The obtained ratio $R_p$ of the BCS proton gaps from all three Skyrme parametrisations considered herein, 
are then plotted in Figure~\ref{fig:Rp vs condensation energy} as a function of the
proton condensation energy $E_{cond}^p$.

The ratio $R_p$ increases, albeit rather minimally, with $E_{cond}^p$ defined in equation~(\ref{eq: condensation energy}).
A fit of the data to a linear equation yields
\begin{equation}
    R_p = 0.0181 \: E_{cond}^p + 0.781
\end{equation}
where $E_{cond}^p$ is given in MeV.
This equation allows for an estimation of the reduction factor to the M\"{o}ller-Nix parameter for any Skyrme parametrisation
and at a given initial pairing matrix element $V_p^0$.
Multiplying the initial M\"{o}ller-Nix parameter $r = 4.8$ MeV with the reduction factor, one then obtains
the proton pairing gap to be utilized in the estimation of pairing matrix element 
via equation~(\ref{eq: integration of ave density for pairing matrix element}).

\begin{table*}[]
    \caption{The ratio $R_p$ of BCS proton pairing gap between Slater approximation $\Delta_p^{Slater}$ and exact Coulomb exchange 
        $\Delta_p^{exact}$ calculations
        using the initial neutron and proton pairing matrix elements listed in column 4 and 5 respectively.
        The proton condensation energy defined in equation~(\ref{eq: condensation energy}) are given in column 9.}
        \label{tab:slater exact ratio}
    \begin{ruledtabular}
    \begin{tabular}{*{10}c}
    &   $Z$&    $N$&    $A$&    $V_n^0$&  $V_p^0$&  $\Delta_p^{Slater}$&    $\Delta_p^{exact}$&    $R_p$&  $E_{cond}^p$   \\
    \hline
\multirow{19}{*}{SIII}	&	62	&	94	&	156	&	0.1902	&	0.2686	&	1.0449	&	1.1884	&	0.879	&	5.2579	\\
	&	62	&	96	&	158	&	0.1855	&	0.2682	&	0.9930	&	1.1319	&	0.877	&	4.7770	\\
	&	64	&	96	&	160	&	0.1862	&	0.2600	&	1.0430	&	1.1894	&	0.877	&	5.4411	\\
	&	66	&	98	&	164	&	0.1824	&	0.2520	&	1.0326	&	1.1612	&	0.889	&	5.3512	\\
	&	66	&	100	&	166	&	0.1784	&	0.2515	&	1.0061	&	1.1436	&	0.880	&	5.1995	\\
	&	66	&	102	&	168	&	0.1742	&	0.2511	&	0.9788	&	1.1254	&	0.870	&	5.0449	\\
	&	68	&	104	&	172	&	0.1712	&	0.2433	&	0.9818	&	1.1108	&	0.884	&	5.0721	\\
	&	70	&	104	&	174	&	0.1717	&	0.2363	&	0.9235	&	1.0772	&	0.857	&	4.9115	\\
	&	70	&	106	&	176	&	0.1681	&	0.2358	&	0.8414	&	1.0169	&	0.827	&	4.3853	\\
	&	72	&	106	&	178	&	0.1625	&	0.2291	&	0.9270	&	1.0700	&	0.866	&	4.9968	\\
	&	74	&	108	&	182	&	0.1652	&	0.2227	&	0.9229	&	1.0715	&	0.861	&	5.1554	\\
	&	92	&	142	&	234	&	0.1235	&	0.1765	&	0.9011	&	1.0114	&	0.891	&	5.7972	\\
	&	92	&	148	&	240	&	0.1182	&	0.1756	&	0.9965	&	1.0954	&	0.910	&	6.8322	\\
	&	94	&	148	&	242	&	0.1183	&	0.1720	&	0.8597	&	0.9952	&	0.864	&	5.7574	\\
	&	94	&	150	&	244	&	0.1167	&	0.1718	&	0.9058	&	1.0375	&	0.873	&	6.2669	\\
	&	96	&	148	&	244	&	0.1185	&	0.1686	&	0.7969	&	0.9611	&	0.829	&	5.4776	\\
	&	98	&	154	&	252	&	0.1138	&	0.1646	&	0.9361	&	1.0406	&	0.900	&	6.5775	\\
	&	100	&	154	&	254	&	0.1139	&	0.1615	&	0.8723	&	0.9700	&	0.899	&	5.8271	\\
	&	100	&	156	&	256	&	0.1124	&	0.1612	&	0.8765	&	0.9806	&	0.894	&	5.9661	\\
 \colrule
\multirow{14}{*}{SLy4}	&	62	&	94	&	156	&	0.2013	&	0.2688	&	1.2149	&	1.1111	&	0.915	&	5.4913	\\
	&	62	&	96	&	158	&	0.1964	&	0.2678	&	1.2015	&	1.0971	&	0.913	&	5.3897	\\
	&	64	&	96	&	160	&	0.1970	&	0.2603	&	1.2320	&	1.1356	&	0.922	&	5.8315	\\
	&	66	&	98	&	164	&	0.1931	&	0.2523	&	1.1568	&	1.0111	&	0.874	&	5.3031	\\
	&	66	&	100	&	166	&	0.1889	&	0.2516	&	1.1287	&	0.9786	&	0.867	&	5.0636	\\
	&	66	&	102	&	168	&	0.1847	&	0.2507	&	1.0978	&	0.9427	&	0.859	&	4.8064	\\
	&	68	&	104	&	172	&	0.1815	&	0.2432	&	1.0554	&	0.9247	&	0.876	&	4.5793	\\
	&	72	&	106	&	178	&	0.1789	&	0.2300	&	1.1540	&	1.0573	&	0.916	&	5.7891	\\
	&	74	&	108	&	182	&	0.1753	&	0.2237	&	1.1912	&	1.0646	&	0.894	&	6.3422	\\
	&	92	&	148	&	240	&	0.1263	&	0.1742	&	1.0504	&	0.9242	&	0.880	&	6.3322	\\
	&	94	&	146	&	240	&	0.1283	&	0.1717	&	0.9364	&	0.8173	&	0.873	&	5.1070	\\
	&	94	&	148	&	242	&	0.1265	&	0.1712	&	0.9784	&	0.8606	&	0.880	&	5.5929	\\
	&	94	&	150	&	244	&	0.1248	&	0.1707	&	1.0267	&	0.9148	&	0.891	&	6.1745	\\
	&	96	&	148	&	244	&	0.1267	&	0.1681	&	0.8995	&	0.7800	&	0.867	&	4.8136	\\
	&	98	&	154	&	252	&	0.1216	&	0.1637	&	0.9353	&	0.7838	&	0.838	&	5.3446	\\
	&	100	&	154	&	254	&	0.1217	&	0.1609	&	0.9020	&	0.8601	&	0.954	&	5.0579	\\
	&	100	&	156	&	256	&	0.1201	&	0.1603	&	0.9253	&	0.8174	&	0.883	&	5.3397	\\
 \colrule
\multirow{17}{*}{SkM*}	&	62	&	94	&	156	&	0.1770	&	0.2574	&	1.2016	&	1.0758	&	0.895	&	5.6088	\\
	&	62	&	96	&	158	&	0.1726	&	0.2570	&	1.1636	&	1.0319	&	0.887	&	5.2677	\\
	&	64	&	96	&	160	&	0.1735	&	0.2493	&	1.2251	&	1.1169	&	0.912	&	6.0193	\\
	&	66	&	98	&	164	&	0.1701	&	0.2414	&	1.1526	&	1.0119	&	0.878	&	5.5037	\\
	&	66	&	100	&	166	&	0.1661	&	0.2409	&	1.1356	&	0.9898	&	0.872	&	5.3521	\\
	&	66	&	102	&	168	&	0.1623	&	0.2405	&	1.1239	&	0.9740	&	0.867	&	5.2519	\\
	&	68	&	104	&	172	&	0.1594	&	0.2331	&	1.0271	&	0.9068	&	0.883	&	4.5248	\\
	&	72	&	106	&	178	&	0.1572	&	0.2199	&	1.1150	&	1.0251	&	0.919	&	5.6524	\\
	&	74	&	108	&	182	&	0.1543	&	0.2138	&	1.1472	&	1.0296	&	0.897	&	6.1558	\\
	&	92	&	148	&	240	&	0.1100	&	0.1675	&	1.0164	&	0.8614	&	0.848	&	6.1661	\\
	&	94	&	146	&	240	&	0.1120	&	0.1644	&	0.9456	&	0.8330	&	0.881	&	5.4395	\\
	&	94	&	148	&	242	&	0.1103	&	0.1642	&	0.9824	&	0.8719	&	0.888	&	5.8792	\\
	&	94	&	150	&	244	&	0.1086	&	0.1639	&	1.0222	&	0.9149	&	0.895	&	6.3749	\\
	&	96	&	148	&	244	&	0.1106	&	0.1609	&	0.9179	&	0.8042	&	0.876	&	5.2362	\\
	&	98	&	154	&	252	&	0.1060	&	0.1570	&	0.9236	&	0.7692	&	0.833	&	5.4339	\\
	&	100	&	154	&	254	&	0.1063	&	0.1539	&	0.8662	&	0.7278	&	0.840	&	4.8742	\\
	&	100	&	156	&	256	&	0.1047	&	0.1537	&	0.9012	&	0.7656	&	0.850	&	5.2850	\\
    \end{tabular}
    \end{ruledtabular}
\end{table*}

Having at hand the function $R_p(E^p_{cond})$, we will proceed as follows. 
We perform HF+BCS calculations with Coulomb \textit{\`{a} la} Slater with the matrix elements $V_p^0$. 
We then obtain BCS pairing gaps $\Delta_q$ and in particular the proton gap $\Delta_p (V_p^0) $
which combined with $V_p^0$ will provide us with $E_{cond}^p (V_p^0)$ 
and the associated value $R_p^0$ of the gap ratio defined in eq.~(\ref{eq: R_p}). 
It is clear that this ratio depends on the retained value for the initial proton pairing matrix element $ V_p^0$. 
The necessity for an iterative determination of the correct ratio $R_p$ consistent 
with a corresponding matrix element $ V_p$ should, 
in principle, be advocated. 

However as it will be discussed at length in Appendix~\ref{sec:appendix a}, 
the convergence of this iterative process is indeed very fast. 
Furthermore, upon making some very limited preliminary studies, it is easy to determine \textit{a priori}, 
for a given particle-hole interaction, an interval of initial values of $V_p^0$, 
previously dubbed as \textit{reasonable}, 
such that the particular choices which are made for those, lead to insignificant corrections. 
It appears that the above mentioned choice of $G_n = G_p = 19$ MeV is convenient in this respect. 
We therefore stick to the corresponding initial values of
\begin{equation}
    r_{corr} = r \: R_p^0
    \label{eq:corrected r values initial}
\end{equation}
without reiterative procedure.
We get the final value $V_p^1$ of the proton average pairing matrix element as done in the first stage of
this calculation for a given nucleus and a given interaction, using now the 
corrected value $r_{corr}$ of the M\"{o}ller-Nix parameter for the estimation process.

\section{Technical details}
\label{sec:technical details}

\subsection{Generation of sp levels}
\label{sec:Generation of sp levels}
We have performed HF+BCS calculations for the deformed nuclear ground-states using three Skyrme parametrisations, 
namely the SIII \cite{SIII}, SkM* \cite{SkMs} and SLy4 \cite{SLy4}, for the strong interaction part of the particle-hole interaction. 

The canonical basis is determined upon solving the HF equations 
resulting from the corresponding energy density functional of the one-body reduced density matrix 
including self-consistently the BCS occupation probabilities. 
The eigensolutions of the corresponding one-body Hamiltonian are obtained by projection of their eigenstates 
onto the eigenstates of an axially symmetrical harmonic oscillator, 
a choice consistent with the axial and intrinsic parity symmetries imposed onto our solutions.

The size of the deformation-dependent basis corresponds for spherical solutions to 17 major shells 
(i.e. with $N_0 = 16$ in the notation of \cite{NPA203}). 
The two parameters defining the size and the ellipsoidal deformation of the harmonic oscillator potential 
(i.e. $b$ and $q$ respectively in the notation of \cite{NPA203}) 
are optimised for each nucleus to yield the lowest equilibrium energy. 
Integrals involving the densities are performed using the Gauss-Hermite and Gauss-Laguerre approximate 
integration methods with 50 and 16 mesh points, respectively.

Pairing correlations are only considered in the isospin $ T = 1 $ channel, 
which amounts in practice, 
for the considered nuclei far enough from the $N = Z$ line,
to restrict to neutron-neutron and proton-proton pairing (thus for $\vert T_z\vert = 1$).   
As already mentioned in Section~\ref{sec:intro} we define this residual pairing interaction $\widehat{v}_{res}$
from an average of its BCS matrix elements
\begin{equation}
    V_q^{ij} = \langle{i_q \bar{i_q}}\vert \widehat{v}_{res} \Big( \vert{j_q \bar{j_q}\rangle - |\bar{j}_q j_q} \rangle \Big) 
\end{equation}
for a given pair $(i,j)$ of the canonical basis states.

Now, we make a further phenomenological step considering as in Ref. \cite{Bonche_pairing} 
a specific dependence of $V_q$  on the neutron or proton numbers $N_q$ in the following form 
\begin{equation}
    V_q = \frac{G_q}{11 + N_q} ,
    \label{eq: seniority force}
\end{equation}
and in what follows $G_q$ will be referred to as the pairing strength.

The validity of this parametrisation has been demonstrated by the quality of the description 
at the same time of odd-even mass differences and of moments of inertia obtained in Ref. \cite{Hafiza2019}. 
A residual interaction is by definition dependent on the number of fermions (through the dependence of the mean field). 
Yet, it is worth noting that the above parametrisation does not necessarily represent only such a dependence but also, 
and may be primarily, in an average fashion, the corresponding dependence of the sp wavefunctions (e.g. through their size or compactness).

Thus our task here is to determine, as sketched in Section~\ref{sec: approach}, 
the two parameters $G_n$ and $G_p$ (and thus $V_n$ and $V_p$) 
for each of the 35 well and rigidly deformed nuclei in the
rare earth and actinide regions (see Section~\ref{sec:results} for details). 
Our approach depends only on sp spectra 
(through the Fermi levels $\tilde{\lambda}_q$ and the value of the average sp level density at these energies $\tilde{\lambda}_q$). 
To get the sp spectra, we have considered pairing residual matrix elements defined by $G_n = G_p = 19$ MeV, as already mentioned.

\begin{figure}
    \centering
    \includegraphics[width=0.52\textwidth]{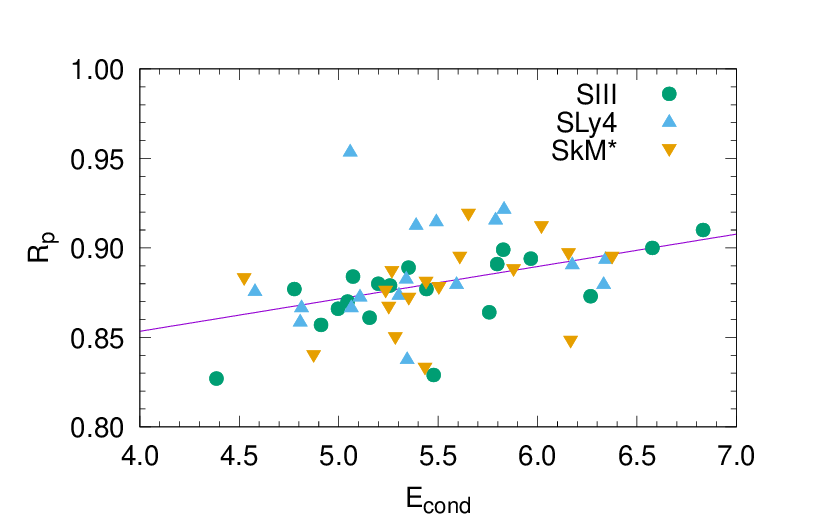}
    \caption{Ratio of the proton BCS gap between exact Coulomb and Slater approximation calculations $R_p$ 
    as a function of the proton pairing condensation energy $E_{cond}$ (given in MeV).}
    \label{fig:Rp vs condensation energy}
\end{figure}

\subsection{Choice of coefficients to determine average level density
\label{sec:choice of coefficients}}

Two important ingredients entering the equation~(\ref{eq: integration of ave density for pairing matrix element}), 
apart from the choice of pairing gap $\tilde{\Delta}_q$
which has been addressed in Section~\ref{sec: effective pairing gaps} 
are the order $M$ of the generalized Laguerre polynomial and the constant $\beta$.

\begin{figure*}
    \centering
    \includegraphics[width=\textwidth]{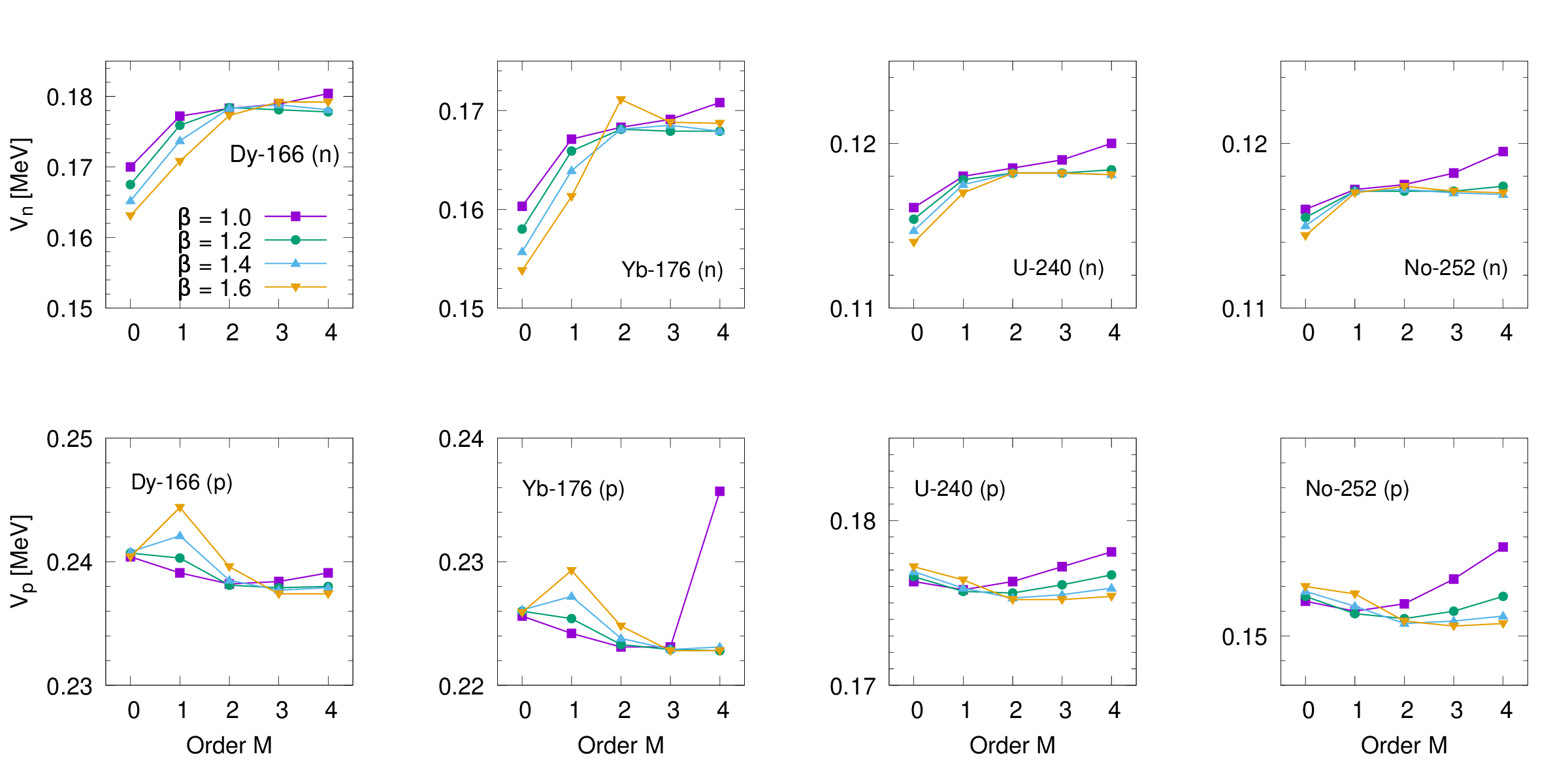}
    \caption{Variation of neutron (top panels) and proton (bottom panels) pairing matrix elements
            showing their evolution as a function of the order $M$ for different values of $\beta$.}
    \label{fig:effect of order and M combined}
\end{figure*}

To determine optimal values of $M$ and $\beta$, we performed fits of pairing matrix elements with $\Omega = 6$ MeV for 
two rare earth ($^{166}$Dy and $^{176}$Yb) and two actinide ($^{240}$U and $^{252}$No) nuclei.
Figure~\ref{fig:effect of order and M combined} shows the variation of neutron (top panels) and proton (bottom panels) 
pairing matrix elements with $M$ and $\beta$ for these nuclei.

\begin{figure*}
    \centering
    \includegraphics[width=\textwidth]{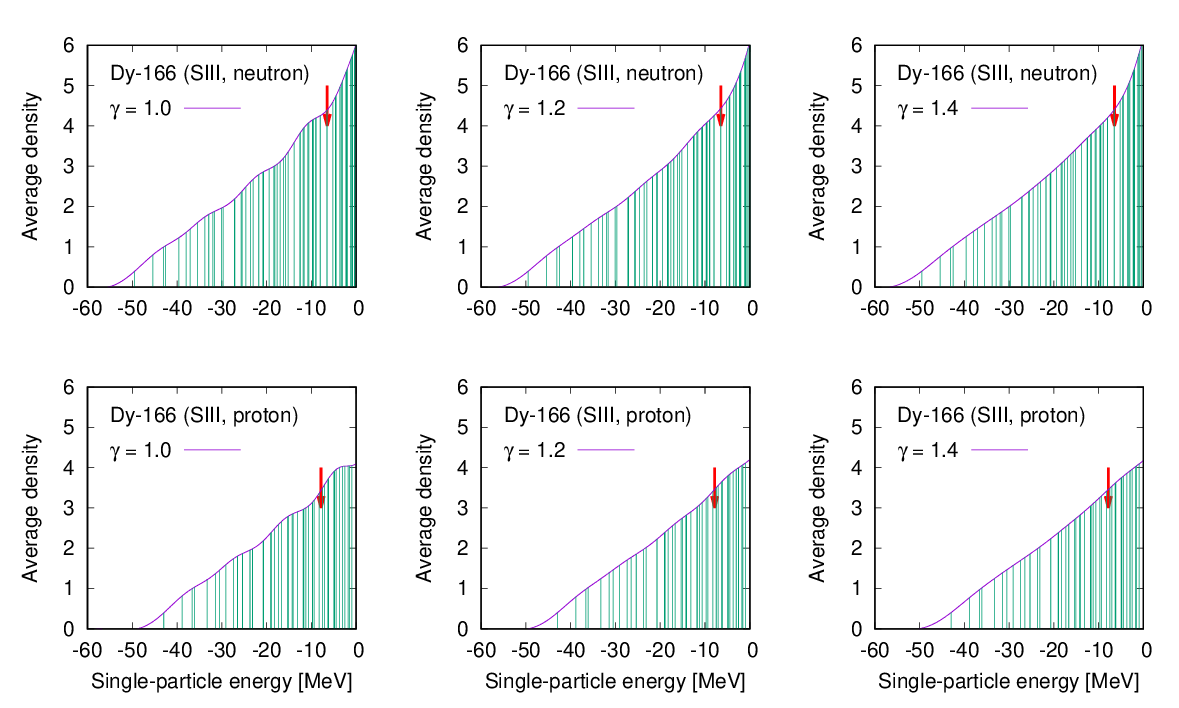}
    \caption{Average neutron (top panels) and proton (bottom panels) 
            sp level densities (in MeV$^{-1}$) using equation~(\ref{eq: integration of ave density for pairing matrix element})
            for the Dy-166 nucleus with $M = 2$ and $\beta = 1.0, 1.2$ and $1.4$ as a function of
            sp energies. Red arrows indicated the location of the Fermi level.}
    \label{fig:average level density Dy-166}
\end{figure*}

We searched for the values of $(M, \beta)$ pairs where a plateau in the average $V_q$ matrix elements is roughly
achieved to ensure that they remain almost constant upon varying $\beta$.
A value of $\beta < 1$ is not sufficient to smooth the shell effect 
(see Figure~\ref{fig:average level density Dy-166})
where some remnants shell effects are still apparent even for $\beta = 1$.
On the other hand, one must avoid too large $\beta$ values (e.g. such that $\beta \ge 1.6$)
to avoid the dubious contribution of unbound sp states poorly
approximated by their projection onto a truncated harmonic oscillator
basis.

From the results displayed in Figure~\ref{fig:effect of order and M combined}, 
we have taken as an optimal choice the following values of the smoothing parameters: $M = 2$ and $\beta = 1.2$.

\section{Results}
\label{sec:results}
Using the approach discussed above, we calculated the MoI using the Inglis-Belyaev \cite{Belyaev} formula
with the estimated pairing strengths unique to each nucleus.
To account approximately for the Thouless-Valatin selfconsistency correction (see Ref.~\cite{Thouless_Valatin})
the calculated MoI have been multiplied by a factor $1.32$,
as suggested in Ref.~\cite{Libert_moi_factor} and shown in previous calculations (see e.g. Ref.~\cite{Hafiza2019})
to provide good estimates of this effect,
for the three Skyrme parametrisations.
The experimental MoI $\mathcal{J}_{exp}$ are determined from the energies
of the first $2^+$ excited state 
in the pure rotor limit (these energies are taken from the compilation of Ref.~\cite{NNDC} and tabulated 
in Table~\ref{tab:moi and fitted pairing results}).

Prior to comparing our calculated MoI with experimental data,
we further eliminate some actinide nuclei which exhibit deficiencies in the sp levels spectra.
These nuclei marked with dashed lines in Table~\ref{tab:moi and fitted pairing results}, all of which are in the heavy nuclei region, shows 
large energy gap located at incorrect nucleon number.
In such nuclei, comparing the calculated MoI with experimental data
would not be meaningful as the observed deviation is not due to the estimation procedure proposed herein,
but rather due to the underlying mean-field solution,
providing locally an inadequate sp level distribution at the Fermi surface.

Indeed, the proposed method which relies on an semi-classical averaging 
is \textit{blind} to the existence of a large sp energy gap.
As examples to this point, we refer to the $^{246}$Pu and $^{248}$Cm ($N = 152$) isotones 
listed in Table~\ref{tab:moi and fitted pairing results}.
In Ref. \cite{Quentin_Bulg_2021} it was reported that an incorrect energy gap was found at $N = 152$ with the SIII parametrisation.
This deficiency is not propagated to the estimated $V_n$ as can be clearly seen 
when comparing the estimated $V_n$ with its neighbouring nuclei.

For comparison between our calculated values with experimental MoI, 
we look at the root-mean-square (r.m.s) deviation $\chi_{\mathcal{J}}$ such that
\begin{equation}
    \chi_{\mathcal{J}} = \sqrt{\frac{\sum_i^N (\mathcal{J}_{TV}^i - \mathcal{J}_{exp}^i)^2 }{N}}
    \label{eq: rms moi}
\end{equation}
where $N$ is the total number of sample nuclei.
The r.m.s deviation have been analysed 
for 23 nuclei around rare-earth region with all three Skyrme parametrisations
and 12 (13) actinide nuclei with SIII and SLy4 (SkM*)
separately.
%, and a combination of both regions with a total of 35 to 36 nuclei.
The $\chi_{\mathcal{J}}$ are tabulated in the top part of Table~\ref{tab:rms moi}.

We found that the $\chi_{\mathcal{J}}$ values ranges from $1.7$ to $3.0$ $\hbar^2/\mbox{MeV}$ in the rare-earth region
and from $3.4$ to $4.6$ $\hbar^2/\mbox{MeV}$ in the actinide region.
It is interesting to note here that the value of $\chi_{\mathcal{J}} \sim 1.77 \: \hbar^2/\mbox{MeV}$ 
for rare earth nuclei only
is indeed very close to the r.m.s deviation ($1.75 \: \hbar^2/\mbox{MeV}$) obtained from average pairing strengths 
fitted to experimental MoI in Ref.~\cite{Hafiza2019}.

In comparing the $\chi_{\mathcal{J}}$ values, it appears that the agreement with data is less spectacular 
in the actinide region as compared to the rare-earth region.
However, comparison of r.m.s deviation based on $\chi_{\mathcal{J}}$ defined in equation (\ref{eq: rms moi})
is not suitable since the number $A$ of nucleons are starkly different 
affecting thus the values of MoI through their $A^{5/3}$ dependence.
To remove the $A$ dependence, we compare instead the weighted r.m.s deviation defined as
\begin{equation}
    \chi_{\mathcal{J}}^{A} = \sqrt{\frac{\sum_i^N \Big(\frac{\mathcal{J}_{TV}^i - \mathcal{J}_{exp}^i}{A^{5/3}_i} \Big)^2}{N}}
    \label{eq: weighted rms mass}
\end{equation}
\begin{equation}
    \chi_{\mathcal{J}}^{exp} = \sqrt{\frac{\sum_i^N \Big(\frac{\mathcal{J}_{TV}^i - \mathcal{J}_{exp}^i}{\mathcal{J}_{exp}^i} \Big)^2}{N}}.
    \label{eq: weighted rms moi}
\end{equation}
In doing so, we see that the agreement with experimental MoI 
are indeed better for actinides than rare-earth region for the SkM* and SLy4 parametrisations
(see Table~\ref{tab:rms moi}).

Finally, we show that these values of the weighted r.m.s deviation provides a way to estimate the uncertainty in
the calculated MoI uniquely for each nucleus.
We define two uncertainty ranges for $\mathcal{J}$ associated to a given nucleus displayed 
in Table~\ref{tab:uncertainty in MoI per nucleus} as $\Delta_\mathcal{J}^A$ (resp. $ \Delta_\mathcal{J}^{exp}$) 
by multiplying $\chi_\mathcal{J}^A$  (resp. $\chi_\mathcal{J}^{exp}$ ) by $A^{5/3}$ (resp. by $J_{TV}$).

\begin{table}
\renewcommand*{\arraystretch}{1.5}
\caption{Weighted root-mean-square deviations $\chi_{\mathcal{J}}$ (in units of $\hbar^2/\mbox{MeV}$), 
    $10^{4} \: \chi_{\mathcal{J}}^{A}$ (in units of $\hbar^2/\mbox{MeV}$) and  
    $10^{2} \: \chi_{\mathcal{J}}^{exp}$
    as defined in equations (\ref{eq: rms moi}), (\ref{eq: weighted rms mass}) and (\ref{eq: weighted rms moi}
    between calculated and experimentally defined 
    MoI obtained with the three Skryme parametrisations.}
\label{tab:rms moi}
    \begin{ruledtabular}
    \begin{tabular}{M{0.08\textwidth}M{0.1\textwidth}M{0.08\textwidth}M{0.08\textwidth}M{0.08\textwidth}}
	       &&  SIII&	   SkM*&	   SLy4  \\
    \hline
    \multirow{2}{*}{$\chi_{\mathcal{J}}$}&    Rare earth&	  1.769&	2.706&	2.940   \\
        &   Actinide&	  4.582&	3.394&	3.498 \\
%        &   Total&	3.042&	2.973&	3.142  \\
    \hline
    \multirow{2}{*}{$\chi_{\mathcal{J}}^{A}$}&    Rare earth&	3.515&    5.080&  5.883   \\
        &   Actinide&	4.844&    3.588&  3.621   \\
%        &   Total&	4.021&   4.597&  5.219   \\
	\hline
    \multirow{2}{*}{$\chi_{\mathcal{J}}^{exp}$}&    Rare earth&	4.548&	7.230&	7.551  \\
        &   Actinide&   6.671&	5.168&   	5.248    \\
%        &   Total&  5.371&	6.560&	6.849  \\
    \end{tabular}
    \end{ruledtabular}   
\end{table}

\begin{table}
    \centering
    \caption{Uncertainty ranges for the moment of inertia  (in units of $\hbar^2$/MeV)
        of some nuclei $\Delta_J^A$ 
            and $\Delta_J^{exp}$ as defined in the text.
    }
    \label{tab:uncertainty in MoI per nucleus}
    \begin{ruledtabular}
        \begin{tabular}{*{6}c}
         Z&     N&      $\mathcal{J}_{TV}$&     $\mathcal{J}_{exp}$&    
                $\Delta^{A}_{\mathcal{J}}$&   $\Delta^{exp}_{\mathcal{J}}$ \\
         \colrule
         62&    98&     40.765&     42.373&     1.658&      1.854   \\
         68&    100&    36.297&     37.592&     1.798&      1.651   \\
         72&    106&    31.988&     32.196&     1.980&      1.455   \\
         92&    140&    63.879&     63.061&     4.243&      4.261   \\
         94&    150&    63.423&     67.873&     4.615&      4.231   \\
         102&   150&    67.468&     64.655&     4.870&      4.501   \\
        \end{tabular}
    \end{ruledtabular}
\end{table}

\section{Conclusion}
\label{sec: conclusion}
In this paper, we have proposed and discussed a simple and efficient method to treat pairing correlations within a microscopic 
(non-relativistic) description of the structure of atomic nuclei. 
It takes stock on the fact that the intensity of pairing correlations depends crucially on level densities around the Fermi surface. 
It is suited to approaches where one has a good knowledge of the particle-hole interaction (e.g. of the usual Skyrme type) 
yielding (possibly in a self-consistent manner) the normal density matrix (and its canonical basis). 
Then one searches for a relevant approach to the abnormal density matrix, typically in a HF+BCS framework. 

At present, it is limited to a very simple ansatz, 
namely using constant pairing matrix elements for sp states located in the vicinity of the Fermi surface, 
dubbed as the seniority force pairing treatment. 
Moreover, it is only operative a priori, so far, to describe the ground states of well and rigidly deformed nuclei. 
As a result, it  yields, in a well-defined fashion, 
the pairing average matrix elements suited for a given nuclear ground state and a given particle-hole interaction. 
It is furthermore important to recall that, 
by no means it gives any direct access to a residual interaction 
since its output includes some average information on the wavefunctions of the states around the Fermi surface, 
particularly their spatial extensions.

For limited that it is now, it may serve, however, 
as a basis to determine the ingredients of a more elaborate pairing treatment, 
i.e. defining univocally a residual pairing interaction 
(contingent now merely on the choice of the particle -- hole interaction)
and not some average of its matrix elements. 
This would allow to build up a BCS treatment which can be used more safely on two counts: 
for all nuclei and away from their equilibrium deformation. 
On the one hand, it would include explicitely (and not in the average) information 
on the structure of the relevant sp wavefunctions. 
And defining an effective residual Hamiltonian, it could be used in a natural fashion away 
from the limited region where it has made sense to have it fitted with some piece of data, on another hand. 
The study of this necessary extension is currently under completion and will be presented in a forthcoming publication. 
To reach that goal it was therefore necessary to assess, first, the quality of the approach discussed here, 
for what it has been specifically tailored.

To determine the relevant pairing matrix elements, 
we used smoothly varying (with nucleon numbers) gaps as in Refs. \cite{Jensen1984,Madland1988}
corrected according to the prescription of Ref.~\cite{Moller1992}.
Experimental odd-even mass difference $\delta E$ are the raw data from where these gaps are extracted. 
Such energies depend of course on the shell structure which is generally well reproduced by state of the art microscopic calculations, 
yet allowing in some well localised regions to generate some misplacement or bad rendering of sp gap intensities 
in some low sp level density regions. 
This is why the information from existing fits of experimental data in terms of smoothly varying 
(with respect to the nucleon numbers) has not been directly compared with 
what results from quantal calculations but with their underlying semi-classical content, determined  in an approximate fashion. 
The mere ingredient of the latter relevant to pairing properties within our pairing model 
is the average sp level density at the Fermi surface and the nucleon numbers of the considered nucleus.

In doing so two important features, absent so far in fits of the pairing intensity, to the best of our knowledge, 
have been carefully taken into account. One is the correction advocated by M\"{o}ller and Nix \cite{Moller1992} 
due to the unescapable selection of data corresponding to sp level densities systematically lower than average. 
The second is due to a systematic overestimation of the proton sp level density at the Fermi surface resulting from the local 
Slater approximation of the Coulomb exchange contribution to the total energy \cite{Titin,Skalski,Bloas}.

The test of our method has consisted in using the so-determined average pairing matrix elements 
(with three different Skyrme force parametrisations) to compare within an Inglis Belyaev approach 
(plus some approximate Thouless-Valatin correction) MoI of about forty well and rigidly deformed rare-earth and actinide nuclei, 
with what is deduced from the experimental energies of their first $2^{+}$ levels. 

These sets of Skyrme parametrisations were chosen to allow us
to assess the versatility of our pairing matrix elements estimation approach using
different mean-field solutions.
It was not our intention here to evaluate the quantal deficiency of the sp spectrum produced using
any of the Skyrme parametrisations.
In fact, we showed that the pairing matrix elements estimated using the proposed approach
herein is not affected by the wrong reproduction of the sp energy gap.
Therefore, this deficiency is irrelevant for a test of our approach. 
Consequently, we had to remove these nuclei in our samples because
implementing them in our test would unduly affect the quality assessment of the proposed method.

From the remaining list of \textit{good} sample nuclei, we found a rather good reproduction of the
\textit{experimental} MoI.
This is especially so for the case of the Skyrme SIII parametrisation in the rare earth region
where our present estimates give excellent agreement with results obtained from a direct fit to these MoI.

This gives confidence on the relevance of what is proposed here 
and allows to take stock on it to tackle our more ambitious attempt to define pairing residual interactions 
from averaged $\delta E$ data, using merely  average sp level densities at the Fermi surface of the calculated canonical basis.

\begin{acknowledgments}
This work was finalized during a research visit in France funded by 
the French Embassy in Malaysia
via the \textit{Mobility Programmes to Support French -- Malaysian Cooperation in Research and Higher Education}
which M.H.K is grateful for.
Gratitude also goes to LP2I Bordeaux and IHPC Strasbourg for the warm hospitality 
extended to him during the visit.
M.H.K would also like to acknowlege Universiti Teknologi Malaysia for its
UTMShine grant (grant number Q.J130000.2454.09G96).
\end{acknowledgments}

\appendix
\section{Treatment of the arbitrariness of the initial proton matrix element $V_p^0$}
\label{sec:appendix a}

Here, we discuss how to circumvent the arbitrariness of the chosen initial pairing matrix element $V_p^0$.
Upon calculating the ground state deformation solutions of our selected nuclei within the HF+BCS approach
(using the seniority force model) we get proton pairing condensation energies. 
They are, as already noted,  dependent upon the choice of the average pairing matrix elements in use in these calculations. 
From them, using equation~(\ref{eq:corrected r values initial}) we should define new values of the proton gaps 
from which using the uniform gap method and the sp spectra, 
we could generate new proton pairing matrix elements $V^1_p$. 
And we could iterate this process to get a convergence of the pairing matrix elements.

The question now is how much change from $V_p^0$ could we expect for these new matrix elements $V^1_p$. 
To estimate that, we assume that we have chosen \textit{reasonable} values of the original $V_p^0$, 
i.e. quantified, for instance, by a deviation of the resulting proton $E_{cond}$ from the value obtained 
through a converged solution of this iterative process by no more than $\pm \: 20 \%$. 
We see from Figure~\ref{fig:Rp vs condensation energy} that such an interval for $E_{cond}$ corresponds to an interval for  $R_p$ 
(and thus on the proton gaps used in the fit) of $\sim \pm \: 2 \%$. 
From the approximation (shown to be rather good, see Figure~\ref{fig:difference pairing matrix element exact vs asinh}) 
of equation~\ref{eq: asinh function for pairing matrix element} concerning the relation between $V_p$ and $\Delta_p$, 
we get readily
\begin{equation}
\frac{\delta V_p}{V_p} = \frac{x}{ln(\sqrt{x + \sqrt{x^2 + 1}})} \frac{1}{\sqrt{ x^2 + 1}} \frac{\delta \Delta _p}{\Delta _p }   
\end{equation}
with $x = \Omega / \Delta_p$.
Using the following values $\Omega = 6$ MeV and $\Delta_p = 1$ MeV, 
one obtains an uncertainty on $V_p$ within the $\pm \: 0.8 \% $ range. 

However, a specific convergence study has been performed for the two deformed nuclei considered in 
Table~\ref{tab:compare G_q when making fit} ($^{176}$Yb and $^{240}$Pu) 
for the three values of the proton pairing intensity parameter $G_p = 16, 19, 22$ MeV. 
One sees on Table~\ref{tab: iteration for Vp} that the value of the proton matrix element $V_p$ 
is converged at the keV level at the second or third iteration. 
Similarly, the mass quadrupole moment $Q_{20}$ is converged at the fm$^2$ level already at the third iteration, 
even sometimes at the second. 
The other lesson comparing the starting value of $V^0_p$ and the converged one, 
is that one may hint that a \textit{reasonable} range for $V^0_p$ values would lie in between $G_p = 16$ and $G_p = 19$ MeV. 
This constitutes a very simple preliminary study, for any given specific particle-hole interaction, 
allowing to define a priori for a global study some value of $V^0_p$ close at
a level of about $1\%$ to what an iterative process would produce. 

In view of the rough nature inherent to the averaging character of the relation between $E_{cond}$ and $R_p$ 
we deem that the iterative process sketched above presents no solid practical justification 
to this paper which aims to illustrate a method for determination of pairing strengths with limited dependence on experimental data.

\begin{table*}[]
    \caption{Estimated proton pairing matrix elements $V_p$ (in MeV) at the ground-state quadrupole moments $Q_{20}$ (in barns)
            obtained for $^{176}$Yb and $^{240}$Pu at corresponding iteration number
            with different starting initial pairing strength $G_p^0 = 16, 19, 22$ MeV.
            The estimated $V_p$ of the preceding iteration are used as initial values 
            for subsequent HF+BCS calculations.}
            \label{tab: iteration for Vp}
    \begin{ruledtabular}
        \begin{tabular}{*{8}c}
        \multirow{2}{*}{Nucleus}&    \multirow{2}{*}{Iteration} &   
                    \multicolumn{2}{c}{$G_p^0 = 16$ MeV}&   \multicolumn{2}{c}{$G_p^0 = 19$ MeV}&   
                    \multicolumn{2}{c}{$G_p^0 = 22$ MeV} \\
                    \cline{3-8}
        	&		&	$Q_{20}$&	$V_p$ &	$Q_{20}$&	$V_p$& $Q_{20}$&	$V_p$    \\
        \hline
        \multirow{5}{*}{Yb-176}	&	1	&	19.13	&	0.2151	&	18.73	&	0.2215	&	18.35	&	0.2300	\\
                                &   2	&	18.89	&	0.2179	&	18.80	&	0.2189	&	18.74	&	0.2206	\\
                                &	3	&	18.88	&	0.2183	&	18.85	&	0.2185	&	18.83	&	0.2188	\\
                                &	4	&	18.86	&	0.2184	&	18.86	&	0.2185	&	18.85	&	0.2185	\\
                            	&	5	&	18.86	&	0.2185	&	18.86	&	0.2185	&	18.86	&	0.2185	\\
        \colrule
        \multirow{5}{*}{Pu-240}	&	1	&	28.72	&	0.1602	&	28.26	&	0.1656	&	27.77	&	0.1733	\\
                            	&	2	&	28.35	&	0.1614	&	28.34	&	0.1624	&	28.32	&	0.1639	\\
                            	&	3	&	28.35	&	0.1615	&	28.35	&	0.1617	&	28.35	&	0.1620	\\
                            	&	4	&	28.35	&	0.1616	&	28.31	&	0.1617	&	28.35	&	0.1616	\\
                                &	5	&	28.35	&	0.1616	&	28.35	&	0.1616	&	28.35	&	0.1616	\\
        \end{tabular}
    \end{ruledtabular}
\end{table*}

\bibliography{apssamp}% Produces the bibliography via BibTeX.

\begin{thebibliography}{28}%
\makeatletter
\providecommand \@ifxundefined [1]{%
 \@ifx{#1\undefined}
}%
\providecommand \@ifnum [1]{%
 \ifnum #1\expandafter \@firstoftwo
 \else \expandafter \@secondoftwo
 \fi
}%
\providecommand \@ifx [1]{%
 \ifx #1\expandafter \@firstoftwo
 \else \expandafter \@secondoftwo
 \fi
}%
\providecommand \natexlab [1]{#1}%
\providecommand \enquote  [1]{``#1''}%
\providecommand \bibnamefont  [1]{#1}%
\providecommand \bibfnamefont [1]{#1}%
\providecommand \citenamefont [1]{#1}%
\providecommand \href@noop [0]{\@secondoftwo}%
\providecommand \href [0]{\begingroup \@sanitize@url \@href}%
\providecommand \@href[1]{\@@startlink{#1}\@@href}%
\providecommand \@@href[1]{\endgroup#1\@@endlink}%
\providecommand \@sanitize@url [0]{\catcode `\\12\catcode `\$12\catcode `\&12\catcode `\#12\catcode `\^12\catcode `\_12\catcode `\%12\relax}%
\providecommand \@@startlink[1]{}%
\providecommand \@@endlink[0]{}%
\providecommand \url  [0]{\begingroup\@sanitize@url \@url }%
\providecommand \@url [1]{\endgroup\@href {#1}{\urlprefix }}%
\providecommand \urlprefix  [0]{URL }%
\providecommand \Eprint [0]{\href }%
\providecommand \doibase [0]{https://doi.org/}%
\providecommand \selectlanguage [0]{\@gobble}%
\providecommand \bibinfo  [0]{\@secondoftwo}%
\providecommand \bibfield  [0]{\@secondoftwo}%
\providecommand \translation [1]{[#1]}%
\providecommand \BibitemOpen [0]{}%
\providecommand \bibitemStop [0]{}%
\providecommand \bibitemNoStop [0]{.\EOS\space}%
\providecommand \EOS [0]{\spacefactor3000\relax}%
\providecommand \BibitemShut  [1]{\csname bibitem#1\endcsname}%
\let\auto@bib@innerbib\@empty
%</preamble>
\bibitem [{\citenamefont {Flocard}\ \emph {et~al.}(1973)\citenamefont {Flocard}, \citenamefont {Quentin}, \citenamefont {Kerman},\ and\ \citenamefont {Vautherin}}]{NPA203}%
  \BibitemOpen
  \bibfield  {author} {\bibinfo {author} {\bibfnamefont {H.}~\bibnamefont {Flocard}}, \bibinfo {author} {\bibfnamefont {P.}~\bibnamefont {Quentin}}, \bibinfo {author} {\bibfnamefont {A.}~\bibnamefont {Kerman}},\ and\ \bibinfo {author} {\bibfnamefont {D.}~\bibnamefont {Vautherin}},\ }\href@noop {} {\bibfield  {journal} {\bibinfo  {journal} {Nucl. Phys. A}\ }\textbf {\bibinfo {volume} {203}},\ \bibinfo {pages} {433} (\bibinfo {year} {1973})}\BibitemShut {NoStop}%
\bibitem [{\citenamefont {Pillet}\ \emph {et~al.}(2002)\citenamefont {Pillet}, \citenamefont {Quentin},\ and\ \citenamefont {Libert}}]{HTDA}%
  \BibitemOpen
  \bibfield  {author} {\bibinfo {author} {\bibfnamefont {N.}~\bibnamefont {Pillet}}, \bibinfo {author} {\bibfnamefont {P.}~\bibnamefont {Quentin}},\ and\ \bibinfo {author} {\bibfnamefont {J.}~\bibnamefont {Libert}},\ }\href@noop {} {\bibfield  {journal} {\bibinfo  {journal} {Nucl. Phys. A}\ }\textbf {\bibinfo {volume} {697}},\ \bibinfo {pages} {141} (\bibinfo {year} {2002})}\BibitemShut {NoStop}%
\bibitem [{\citenamefont {Bochnacki}\ \emph {et~al.}(1967)\citenamefont {Bochnacki}, \citenamefont {Holban},\ and\ \citenamefont {Mikhailov}}]{SurfaceDelta}%
  \BibitemOpen
  \bibfield  {author} {\bibinfo {author} {\bibfnamefont {Z.}~\bibnamefont {Bochnacki}}, \bibinfo {author} {\bibfnamefont {I.}~\bibnamefont {Holban}},\ and\ \bibinfo {author} {\bibfnamefont {I.}~\bibnamefont {Mikhailov}},\ }\href@noop {} {\bibfield  {journal} {\bibinfo  {journal} {Nucl. Phys. A}\ }\textbf {\bibinfo {volume} {97}},\ \bibinfo {pages} {33} (\bibinfo {year} {1967})}\BibitemShut {NoStop}%
\bibitem [{\citenamefont {Tian}\ \emph {et~al.}(2009)\citenamefont {Tian}, \citenamefont {Ma},\ and\ \citenamefont {Ring}}]{GaussPsep}%
  \BibitemOpen
  \bibfield  {author} {\bibinfo {author} {\bibfnamefont {Y.}~\bibnamefont {Tian}}, \bibinfo {author} {\bibfnamefont {Z.}~\bibnamefont {Ma}},\ and\ \bibinfo {author} {\bibfnamefont {P.}~\bibnamefont {Ring}},\ }\href@noop {} {\bibfield  {journal} {\bibinfo  {journal} {Phys. Lett. B}\ }\textbf {\bibinfo {volume} {676}},\ \bibinfo {pages} {44} (\bibinfo {year} {2009})}\BibitemShut {NoStop}%
\bibitem [{\citenamefont {Bohr}\ \emph {et~al.}(1958)\citenamefont {Bohr}, \citenamefont {Mottelson},\ and\ \citenamefont {Pines}}]{BMP}%
  \BibitemOpen
  \bibfield  {author} {\bibinfo {author} {\bibfnamefont {A.}~\bibnamefont {Bohr}}, \bibinfo {author} {\bibfnamefont {B.~R.}\ \bibnamefont {Mottelson}},\ and\ \bibinfo {author} {\bibfnamefont {D.}~\bibnamefont {Pines}},\ }\href@noop {} {\bibfield  {journal} {\bibinfo  {journal} {Phys. Rev.}\ }\textbf {\bibinfo {volume} {110}},\ \bibinfo {pages} {936} (\bibinfo {year} {1958})}\BibitemShut {NoStop}%
\bibitem [{\citenamefont {Nor}\ \emph {et~al.}(2019)\citenamefont {Nor}, \citenamefont {Rezle}, \citenamefont {Kelvin-Lee}, \citenamefont {Koh}, \citenamefont {Bonneau},\ and\ \citenamefont {Quentin}}]{Hafiza2019}%
  \BibitemOpen
  \bibfield  {author} {\bibinfo {author} {\bibfnamefont {N.~M.}\ \bibnamefont {Nor}}, \bibinfo {author} {\bibfnamefont {N.-A.}\ \bibnamefont {Rezle}}, \bibinfo {author} {\bibfnamefont {K.-W.}\ \bibnamefont {Kelvin-Lee}}, \bibinfo {author} {\bibfnamefont {M.-H.}\ \bibnamefont {Koh}}, \bibinfo {author} {\bibfnamefont {L.}~\bibnamefont {Bonneau}},\ and\ \bibinfo {author} {\bibfnamefont {P.}~\bibnamefont {Quentin}},\ }\href@noop {} {\bibfield  {journal} {\bibinfo  {journal} {Phys. Rev. C}\ }\textbf {\bibinfo {volume} {99}},\ \bibinfo {pages} {064306} (\bibinfo {year} {2019})}\BibitemShut {NoStop}%
\bibitem [{\citenamefont {Brack}\ \emph {et~al.}(1972)\citenamefont {Brack}, \citenamefont {Damgaard},\ and\ \citenamefont {\textit{et al.}}}]{FunnyHill}%
  \BibitemOpen
  \bibfield  {author} {\bibinfo {author} {\bibfnamefont {M.}~\bibnamefont {Brack}}, \bibinfo {author} {\bibfnamefont {J.}~\bibnamefont {Damgaard}},\ and\ \bibinfo {author} {\bibfnamefont {A.~J.}\ \bibnamefont {\textit{et al.}}},\ }\href@noop {} {\bibfield  {journal} {\bibinfo  {journal} {Rev. Mod. Phys.}\ }\textbf {\bibinfo {volume} {44}},\ \bibinfo {pages} {320} (\bibinfo {year} {1972})}\BibitemShut {NoStop}%
\bibitem [{\citenamefont {Ring}\ and\ \citenamefont {Schuck}(1980)}]{RingSchuck}%
  \BibitemOpen
  \bibfield  {author} {\bibinfo {author} {\bibfnamefont {P.}~\bibnamefont {Ring}}\ and\ \bibinfo {author} {\bibfnamefont {P.}~\bibnamefont {Schuck}},\ }\href@noop {} {\emph {\bibinfo {title} {The Nuclear Many-body Problem}}}\ (\bibinfo  {publisher} {Springer-Verlag},\ \bibinfo {year} {1980})\ p.\ \bibinfo {pages} {240}\BibitemShut {NoStop}%
\bibitem [{\citenamefont {Jensen}\ and\ \citenamefont {Hansen}(1984)}]{Jensen1984}%
  \BibitemOpen
  \bibfield  {author} {\bibinfo {author} {\bibfnamefont {A.~S.}\ \bibnamefont {Jensen}}\ and\ \bibinfo {author} {\bibfnamefont {P.~G.}\ \bibnamefont {Hansen}},\ }\href@noop {} {\bibfield  {journal} {\bibinfo  {journal} {Nucl. Phys. A}\ }\textbf {\bibinfo {volume} {431}},\ \bibinfo {pages} {393} (\bibinfo {year} {1984})}\BibitemShut {NoStop}%
\bibitem [{\citenamefont {Madland}\ and\ \citenamefont {Nix}(1988)}]{Madland1988}%
  \BibitemOpen
  \bibfield  {author} {\bibinfo {author} {\bibfnamefont {D.}~\bibnamefont {Madland}}\ and\ \bibinfo {author} {\bibfnamefont {J.}~\bibnamefont {Nix}},\ }\href@noop {} {\bibfield  {journal} {\bibinfo  {journal} {Nucl. Phys. A}\ }\textbf {\bibinfo {volume} {476}},\ \bibinfo {pages} {1 } (\bibinfo {year} {1988})}\BibitemShut {NoStop}%
\bibitem [{\citenamefont {M{\"{o}}ller}\ and\ \citenamefont {Nix}(1992)}]{Moller1992}%
  \BibitemOpen
  \bibfield  {author} {\bibinfo {author} {\bibfnamefont {P.}~\bibnamefont {M{\"{o}}ller}}\ and\ \bibinfo {author} {\bibfnamefont {J.}~\bibnamefont {Nix}},\ }\href@noop {} {\bibfield  {journal} {\bibinfo  {journal} {Nucl. Phys. A}\ }\textbf {\bibinfo {volume} {536}},\ \bibinfo {pages} {20} (\bibinfo {year} {1992})}\BibitemShut {NoStop}%
\bibitem [{\citenamefont {Slater}(1951)}]{Slater}%
  \BibitemOpen
  \bibfield  {author} {\bibinfo {author} {\bibfnamefont {J.}~\bibnamefont {Slater}},\ }\href@noop {} {\bibfield  {journal} {\bibinfo  {journal} {Phys. Rev.}\ }\textbf {\bibinfo {volume} {81}},\ \bibinfo {pages} {385} (\bibinfo {year} {1951})}\BibitemShut {NoStop}%
\bibitem [{\citenamefont {Titin-Schnaider}\ and\ \citenamefont {Quentin}(1974)}]{Titin}%
  \BibitemOpen
  \bibfield  {author} {\bibinfo {author} {\bibfnamefont {C.}~\bibnamefont {Titin-Schnaider}}\ and\ \bibinfo {author} {\bibfnamefont {P.}~\bibnamefont {Quentin}},\ }\href@noop {} {\bibfield  {journal} {\bibinfo  {journal} {Phys. Lett. B}\ }\textbf {\bibinfo {volume} {49}},\ \bibinfo {pages} {397} (\bibinfo {year} {1974})}\BibitemShut {NoStop}%
\bibitem [{\citenamefont {Skalski}(2001)}]{Skalski}%
  \BibitemOpen
  \bibfield  {author} {\bibinfo {author} {\bibfnamefont {J.}~\bibnamefont {Skalski}},\ }\href@noop {} {\bibfield  {journal} {\bibinfo  {journal} {Phys. Rev. C}\ }\textbf {\bibinfo {volume} {63}},\ \bibinfo {pages} {024312} (\bibinfo {year} {2001})}\BibitemShut {NoStop}%
\bibitem [{\citenamefont {Bloas}\ \emph {et~al.}(2011)\citenamefont {Bloas}, \citenamefont {Koh}, \citenamefont {Quentin}, \citenamefont {Bonneau},\ and\ \citenamefont {Ithnin}}]{Bloas}%
  \BibitemOpen
  \bibfield  {author} {\bibinfo {author} {\bibfnamefont {J.~L.}\ \bibnamefont {Bloas}}, \bibinfo {author} {\bibfnamefont {M.-H.}\ \bibnamefont {Koh}}, \bibinfo {author} {\bibfnamefont {P.}~\bibnamefont {Quentin}}, \bibinfo {author} {\bibfnamefont {L.}~\bibnamefont {Bonneau}},\ and\ \bibinfo {author} {\bibfnamefont {J.}~\bibnamefont {Ithnin}},\ }\href@noop {} {\bibfield  {journal} {\bibinfo  {journal} {Phys. Rev. C}\ }\textbf {\bibinfo {volume} {84}},\ \bibinfo {pages} {0143310} (\bibinfo {year} {2011})}\BibitemShut {NoStop}%
\bibitem [{\citenamefont {NNDC}()}]{NNDC}%
  \BibitemOpen
  \bibfield  {author} {\bibinfo {author} {\bibnamefont {NNDC}},\ }\href@noop {} {\bibinfo  {journal} {\url{https://www.nndc.bnl.gov/nudat3/}}\ }\BibitemShut {NoStop}%
\bibitem [{\citenamefont {Beiner}\ \emph {et~al.}(1975)\citenamefont {Beiner}, \citenamefont {Flocard}, \citenamefont {Giai},\ and\ \citenamefont {Quentin}}]{SIII}%
  \BibitemOpen
\bibfield  {journal} {  }\bibfield  {author} {\bibinfo {author} {\bibfnamefont {M.}~\bibnamefont {Beiner}}, \bibinfo {author} {\bibfnamefont {H.}~\bibnamefont {Flocard}}, \bibinfo {author} {\bibfnamefont {N.~V.}\ \bibnamefont {Giai}},\ and\ \bibinfo {author} {\bibfnamefont {P.}~\bibnamefont {Quentin}},\ }\href@noop {} {\bibfield  {journal} {\bibinfo  {journal} {Nucl. Phys. A}\ }\textbf {\bibinfo {volume} {238}},\ \bibinfo {pages} {29} (\bibinfo {year} {1975})}\BibitemShut {NoStop}%
\bibitem [{\citenamefont {Bartel}\ \emph {et~al.}(1982)\citenamefont {Bartel}, \citenamefont {Quentin}, \citenamefont {Brack}, \citenamefont {Guet},\ and\ \citenamefont {H{\aa}kansson}}]{SkMs}%
  \BibitemOpen
  \bibfield  {author} {\bibinfo {author} {\bibfnamefont {J.}~\bibnamefont {Bartel}}, \bibinfo {author} {\bibfnamefont {P.}~\bibnamefont {Quentin}}, \bibinfo {author} {\bibfnamefont {M.}~\bibnamefont {Brack}}, \bibinfo {author} {\bibfnamefont {C.}~\bibnamefont {Guet}},\ and\ \bibinfo {author} {\bibfnamefont {H.-B.}\ \bibnamefont {H{\aa}kansson}},\ }\href@noop {} {\bibfield  {journal} {\bibinfo  {journal} {Nucl. Phys. A}\ }\textbf {\bibinfo {volume} {386}},\ \bibinfo {pages} {79} (\bibinfo {year} {1982})}\BibitemShut {NoStop}%
\bibitem [{\citenamefont {Chabanat}\ \emph {et~al.}(1988)\citenamefont {Chabanat}, \citenamefont {Bonche}, \citenamefont {Haensel}, \citenamefont {Meyer},\ and\ \citenamefont {Schaeffer}}]{SLy4}%
  \BibitemOpen
  \bibfield  {author} {\bibinfo {author} {\bibfnamefont {E.}~\bibnamefont {Chabanat}}, \bibinfo {author} {\bibfnamefont {P.}~\bibnamefont {Bonche}}, \bibinfo {author} {\bibfnamefont {P.}~\bibnamefont {Haensel}}, \bibinfo {author} {\bibfnamefont {J.}~\bibnamefont {Meyer}},\ and\ \bibinfo {author} {\bibfnamefont {R.}~\bibnamefont {Schaeffer}},\ }\href@noop {} {\bibfield  {journal} {\bibinfo  {journal} {Nucl. Phys. A}\ }\textbf {\bibinfo {volume} {635}},\ \bibinfo {pages} {231} (\bibinfo {year} {1988})}\BibitemShut {NoStop}%
\bibitem [{\citenamefont {and. R.K.~Bhaduri}\ and\ \citenamefont {Brack}(1975)}]{Jennings1975}%
  \BibitemOpen
  \bibfield  {author} {\bibinfo {author} {\bibfnamefont {B.~J.}\ \bibnamefont {and. R.K.~Bhaduri}}\ and\ \bibinfo {author} {\bibfnamefont {M.}~\bibnamefont {Brack}},\ }\href@noop {} {\bibfield  {journal} {\bibinfo  {journal} {Nucl. Phys. A}\ }\textbf {\bibinfo {volume} {253}},\ \bibinfo {pages} {29} (\bibinfo {year} {1975})}\BibitemShut {NoStop}%
\bibitem [{\citenamefont {Brack}\ and\ \citenamefont {Pauli}(1973)}]{NPA207}%
  \BibitemOpen
  \bibfield  {author} {\bibinfo {author} {\bibfnamefont {M.}~\bibnamefont {Brack}}\ and\ \bibinfo {author} {\bibfnamefont {H.}~\bibnamefont {Pauli}},\ }\href@noop {} {\bibfield  {journal} {\bibinfo  {journal} {Nucl. Phys. A}\ }\textbf {\bibinfo {volume} {207}},\ \bibinfo {pages} {401} (\bibinfo {year} {1973})}\BibitemShut {NoStop}%
\bibitem [{\citenamefont {Moszkowski}(1957)}]{Moszkowski}%
  \BibitemOpen
  \bibfield  {author} {\bibinfo {author} {\bibfnamefont {S.~A.}\ \bibnamefont {Moszkowski}},\ }\href@noop {} {\bibfield  {journal} {\bibinfo  {journal} {Handbuch der Physik}\ }\textbf {\bibinfo {volume} {XXXIX}},\ \bibinfo {pages} {469} (\bibinfo {year} {1957})}\BibitemShut {NoStop}%
\bibitem [{\citenamefont {Zheng}\ \emph {et~al.}(1992)\citenamefont {Zheng}, \citenamefont {Sprung},\ and\ \citenamefont {Flocard}}]{Zheng_1992}%
  \BibitemOpen
  \bibfield  {author} {\bibinfo {author} {\bibfnamefont {D.~C.}\ \bibnamefont {Zheng}}, \bibinfo {author} {\bibfnamefont {D.~W.~L.}\ \bibnamefont {Sprung}},\ and\ \bibinfo {author} {\bibfnamefont {H.}~\bibnamefont {Flocard}},\ }\href@noop {} {\bibfield  {journal} {\bibinfo  {journal} {Phys. Rev. C}\ }\textbf {\bibinfo {volume} {46}},\ \bibinfo {pages} {1355} (\bibinfo {year} {1992})}\BibitemShut {NoStop}%
\bibitem [{\citenamefont {Bonche}\ \emph {et~al.}(1985)\citenamefont {Bonche}, \citenamefont {Flocard}, \citenamefont {Heenen}, \citenamefont {Krieger},\ and\ \citenamefont {Weiss}}]{Bonche_pairing}%
  \BibitemOpen
  \bibfield  {author} {\bibinfo {author} {\bibfnamefont {P.}~\bibnamefont {Bonche}}, \bibinfo {author} {\bibfnamefont {H.}~\bibnamefont {Flocard}}, \bibinfo {author} {\bibfnamefont {P.}~\bibnamefont {Heenen}}, \bibinfo {author} {\bibfnamefont {S.}~\bibnamefont {Krieger}},\ and\ \bibinfo {author} {\bibfnamefont {M.}~\bibnamefont {Weiss}},\ }\href@noop {} {\bibfield  {journal} {\bibinfo  {journal} {Nucl. Phys. A}\ }\textbf {\bibinfo {volume} {443}},\ \bibinfo {pages} {39} (\bibinfo {year} {1985})}\BibitemShut {NoStop}%
\bibitem [{\citenamefont {Belyaev}(1961)}]{Belyaev}%
  \BibitemOpen
  \bibfield  {author} {\bibinfo {author} {\bibfnamefont {S.}~\bibnamefont {Belyaev}},\ }\href@noop {} {\bibfield  {journal} {\bibinfo  {journal} {Nucl. Phys. A}\ }\textbf {\bibinfo {volume} {24}},\ \bibinfo {pages} {322} (\bibinfo {year} {1961})}\BibitemShut {NoStop}%
\bibitem [{\citenamefont {Thouless}\ and\ \citenamefont {Valatin}(1962)}]{Thouless_Valatin}%
  \BibitemOpen
  \bibfield  {author} {\bibinfo {author} {\bibfnamefont {D.~J.}\ \bibnamefont {Thouless}}\ and\ \bibinfo {author} {\bibfnamefont {J.~G.}\ \bibnamefont {Valatin}},\ }\href@noop {} {\bibfield  {journal} {\bibinfo  {journal} {Nucl. Phys.}\ }\textbf {\bibinfo {volume} {31}},\ \bibinfo {pages} {211} (\bibinfo {year} {1962})}\BibitemShut {NoStop}%
\bibitem [{\citenamefont {Libert}\ \emph {et~al.}(1999)\citenamefont {Libert}, \citenamefont {Girod},\ and\ \citenamefont {Delaroche}}]{Libert_moi_factor}%
  \BibitemOpen
  \bibfield  {author} {\bibinfo {author} {\bibfnamefont {J.}~\bibnamefont {Libert}}, \bibinfo {author} {\bibfnamefont {M.}~\bibnamefont {Girod}},\ and\ \bibinfo {author} {\bibfnamefont {J.}~\bibnamefont {Delaroche}},\ }\href@noop {} {\bibfield  {journal} {\bibinfo  {journal} {Phys. Rev. C}\ }\textbf {\bibinfo {volume} {60}},\ \bibinfo {pages} {054301} (\bibinfo {year} {1999})}\BibitemShut {NoStop}%
\bibitem [{\citenamefont {Quentin}\ \emph {et~al.}(2021)\citenamefont {Quentin}, \citenamefont {Bonneau}, \citenamefont {Minkov},\ and\ \citenamefont {\textit{et al.}}}]{Quentin_Bulg_2021}%
  \BibitemOpen
  \bibfield  {author} {\bibinfo {author} {\bibfnamefont {P.}~\bibnamefont {Quentin}}, \bibinfo {author} {\bibfnamefont {L.}~\bibnamefont {Bonneau}}, \bibinfo {author} {\bibfnamefont {N.}~\bibnamefont {Minkov}},\ and\ \bibinfo {author} {\bibnamefont {\textit{et al.}}},\ }\href@noop {} {\bibfield  {journal} {\bibinfo  {journal} {Bulg. J. Phys}\ }\textbf {\bibinfo {volume} {48}},\ \bibinfo {pages} {634} (\bibinfo {year} {2021})}\BibitemShut {NoStop}%
\end{thebibliography}%

\end{document}